\documentclass{pinchcrMEC}
\renewcommand{\v}{{\bf v}}
\renewcommand{\r}{{\bf r}}
\newcommand{\p}{{\bf p}}
\newcommand{\bsigma}{{\mbox{\boldmath{$\sigma$}}}}
\renewcommand{\S}{{\mathcal S}}
\usepackage{graphicx}
\title{Complex Fluids: The Physics of Emulsions}
\subtitle{Lecture Notes for les Houches 2012 Summer School on Soft Interfaces}
\author{M. E. Cates}
\affiliation{SUPA, School of Physics and Astronomy, University of Edinburgh, JCMB Kings Buildings, Mayfield Road, Edinburgh EH9 3JZ, Scotland}
\begin{document}
\maketitle
\tableofcontents
\maintext
%
\chapter{Introduction}
Oil and water do not mix. Sometimes, however, we want them to mix and stay mixed for long periods of time, in the form of emulsions. The same applies to many other fluid pairs, A and B (say), including many that are molecularly miscible but only in some temperature range (usually high temperature). Quenching the A/B system initiates phase separation; these lectures describe that process and then various ways to halt it before it is complete, so that the resulting fluid domains have a finite length scale. 

Such finely divided mixtures, generally known as emulsions, have many applications in science and technology. Macro-emulsions, containing micron-sized or larger spherical droplets of oil in water or vice versa, are found in many products ranging from foods to agrochemicals. As we shall see these are generally not thermodynamically stable, but various tricks are available to prevent phase separation over long periods of time. Examples of such tricks include use of adsorbed surfactants to inhibit coalescence, or incorporation of trapped species to inhibit diffusive coarsening. 

For fluid pairs of similar phase volumes and viscosities, phase separation in 3D creates bicontinuous structure in which A-rich and B-rich phases each form connected interpenetrating domains. Stabilizing such a structure on a finite length-scale is more difficult than for droplets, although in principle this can be done by subjecting the system to continuous shear. A more practical alternative is to add a well chosen surfactant, capable of reducing the interfacial tension effectively to zero. The resulting states, known as micro-emulsions, are thermodynamically stable and include not only droplet morphologies but bicontinuous ones. Bicontinuous micro-emulsions are fluids containing an interface whose fluctuating random geometry is sustained by entropy. 

Traditional surfactants adsorb and desorb easily and reversibly at the fluid-fluid interface, so that the interface in surfactant-stabilized micro- or macro-emulsions is generally in local equilibrium. This ceases to be the case for much larger amphiphilic objects, such as block copolymers, globular proteins, and janus particles, where interfacial detachment energies can be hundreds or thousands of times $k_BT$. Here the term 'janus beads' refers to spherical colloids with hemispheres of opposite surface chemistry; these obviously behave like large, irreversibly adsorbed surfactants. 

Perhaps less obviously, colloidal spheres of uniform surface chemistry can also become trapped, with similarly large attachment energies, at the fluid-fluid interface. Such structures have maximum local stability when the surface chemistry creates nearly equal interfacial tension between the solid and both fluids (neutral wetting). Stabilization by particles creates interesting alternative avenues to the formation of long-lived emulsified states. Such avenues were long neglected, possibly because to achieve them reproducibly requires detailed control of the entire preparation history of the sample. However this feature, which stems directly from the fact that detachment from the interface cannot be achieved by Brownian motion, is now seen as an opportunity for robustly locking in a desired microstructure. Such locked-in structures can include foams, multiple emulsions of nested droplets, and assemblies of highly nonspherical droplets. They also include bicontinuous states in which the jammed layer imparts rigidity to the entire 3D sample.

In what follows, I will elaborate on the above narrative at the level of detail achievable in a set of four 90 minute lectures. This involves ruthless simplification, sometimes to the threshold of dishonesty (a necessity familiar to physics lecturers everywhere) which I hope I have not overstepped. Several calculations are left as exercises. (Also in some of these I have not carefully checked the numerical prefactors to the answers given -- that forms part of the exercise!) Many important topics are deliberately avoided because they are not central to the story: this applies particularly to systems where A and B are not simple fluids but themselves complex (polymers, suspensions, gels etc.).

I have deliberately kept references through most of the text to the minimum level consistent with clarity. The exception is Section \ref{PE} on particle-stabilized emulsions which describes some relatively recent work. This area apart, there are good graduate texts and monographs that cover between them most of the topics addressed here at varying levels of detail; these include, but are not limited to, References \cite{chaikin,onuki,bibette,safran,nelson,binks,larson,book}.

\chapter{Binary Fluid Phase Separation}
We start by recalling the description of an isothermal, incompressible, simple fluid with newtonian viscosity $\eta$ and density $\rho$. This obeys the Navier Stokes equation (NSE)
\begin{equation}
\rho(\dot \v + \v.\nabla\v) = \eta\nabla^2\v - \nabla P \label{NSE}
\end{equation}
Here the pressure field $P$ must be chosen to enforce the incompressibility condition 
\begin{equation}
\nabla.\v = 0 \label{continuity}
\end{equation}
One generic approach to complex fluids is to consider a simple fluid obeying the NSE, coupled to a set of mesoscopic internal structural variables $X(\r)$, each a function of position. In principle the density and viscosity in (\ref{NSE}) could depend directly on $X(\r)$. However a big simplification, which does not affect much the qualitative physics discussed below, is to assume these dependences are negligible. The main effect of $X(\r)$ is then to create an additional stress tensor that enters the NSE alongside the familiar viscous term:
\begin{equation}
\rho(\dot \v + \v.\nabla\v) = \eta\nabla^2\v - \nabla P 
+\nabla.\bsigma[X(\r)]
\label{NSE2}
\end{equation}
The stress term can alternatively be viewed as a force density ${\bf f} = \nabla\bsigma$ exerted by the structural degrees of freedom on the fluid continuum. 

This stress term can be large, and after suitable averaging its macroscopic effect in many complex fluids includes a greatly increased viscosity at small steady shear rates, and nonlinear rheological phenomena at larger ones. This applies for instance in both polymer solutions (where $X$ describes some coarse-grained conformational variables for the polymers) and nematic liquid crystals (where $X$ stands for the order parameter tensor $Q_{\alpha\beta}$ describing local molecular alignment). For both these systems, a good numerical approach is to proceed via (\ref{NSE2}), supplemented by a recipe for $\bsigma[X(\r)]$ and a time evolution equation for $X$ itself. This approach means that a large body of numerical expertise in solving the forced NSE can be exploited. 

In these lectures we will be concerned with binary mixtures of simple Newtonian fluids, for which the relevant mesoscopic variable is a scalar describing the local composition of the fluid mixture. We define it as
\begin{equation}
\phi(\r) = \frac{\langle n_A-n_B \rangle_{\rm meso}}{\bar{n}_A+\bar{n}_B}
\label{phidef}
\end{equation}
Here $n_{A,B}$ denotes the number of A,B molecules per unit volume locally and the overbar denotes the macroscopic average over the whole system. The mesoscopic average is taken over a large enough (but still small) local volume so that $\phi(\r)$ is a smooth field. For simplicity we have assumed that A and B molecules have the same molecular volume; our assumption that $\rho$ does not depend on $\phi$ then requires that they also have equal mass. Given the incompressibility condition implicit in (\ref{continuity}), we see that our composition variable obeys $-1\le\phi\le 1$ with $\phi=1$ in a fluid of pure A. 

Later on we will show that for such a fluid mixture, 
\begin{equation}
\nabla.\bsigma(\r) = -\phi(\r)\nabla \mu(\r) \equiv -\phi(\r)\nabla\frac{\delta F}{\delta\phi(\r)}
\label{body}
\end{equation}
where $F[\phi]$ is a free energy functional (described further below) and its functional derivative $\mu$ is a chemical potential conjugate to the composition $\phi$. This is properly called the `exchange chemical potential' as it controls the free energy increment on swapping B molecules for A (hence changing $\phi$) at fixed total density.

\section{The symmetric binary fluid}
We now restrict ourselves further to the case where the molecular A-A and B-B interactions are the same but there is an additional repulsion $E_{AB}$ between adjacent molecules of A and B. Combined with the previous assumptions about molecular mass and size, the system is now completely symmetric at a molecular level.  At high temperatures, $T>T_C\simeq E_{AB}/k_B$, the repulsive interactions are overcome by mixing entropy and the two fluids remain completely miscible. At lower temperatures however, the A-B repulsion will cause demixing into two phases, one rich in A, one rich in B. Entropy ensures that there is always a small amount of the other type of molecule present; close to the critical temperature $T_C$ the two phases differ only slightly in $\phi$. 

\begin{figure}[hbtp]
	\centering
	\includegraphics[width=9.truecm]{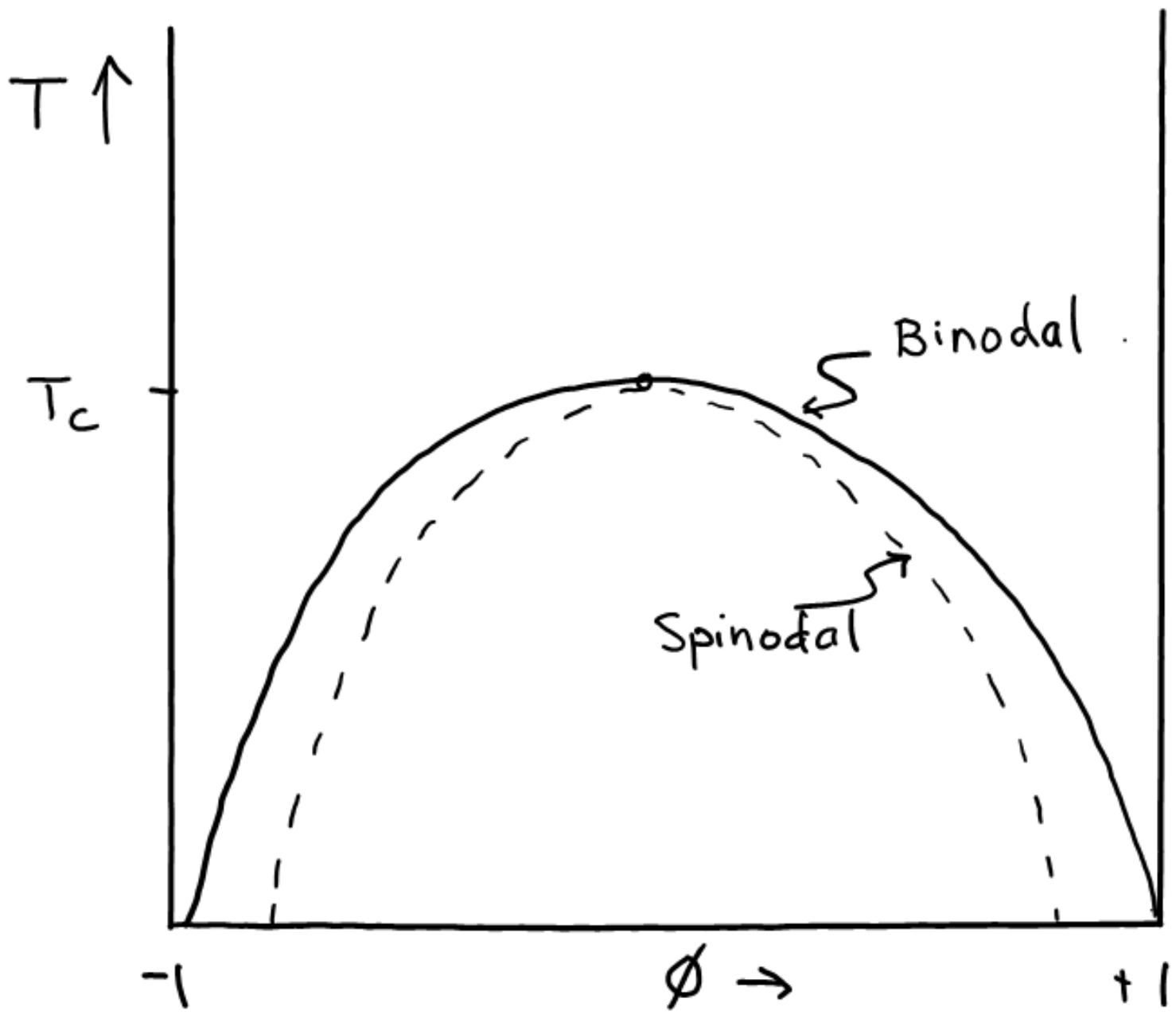}
	\caption[]{Phase diagram of a symmetric binary fluid mixture.}
	\label{one}
\end{figure}

A schematic phase diagram for this system is shown in Fig.\ref{one}.
The locus of co-existing compositions $\phi = \pm \phi_b(T)$ is called the binodal curve; within the binodal, the equilibrium state comprises two coexisting phases of composition $\pm\phi_b$. The amount of each phase depends on the global composition $\bar\phi$ of the initial mixture. Specifically, the volumes occupied by the A-rich and B-rich phases, denoted $\Phi_{A,B}V$, where $V$ is the overall volume of the system, obey $\Phi_A+\Phi_B=1$ (clearly) and
\begin{equation}
(\Phi_A-\Phi_B)\phi_b = \bar\phi
\label{phasevol}
\end{equation}
Thus the phase volume $\Phi_A$ evolves from zero to one as the overall composition $\phi$ is swept across the miscibility gap from $-\phi_b$ to $\phi_b$. 
Note that in what follows, we sometimes refer to the coexisting fluids as simply A and B when the more proper terms would be A-rich phase and B-rich phase. Notationally however we distinguish by capital letters phase volumes $\Phi_{A,B}$ (which lie between zero and unity) from compositions $\phi$ (which obey $-1\le\phi\le 1$). 

The dotted line on the phase diagram is called the spinodal, $\phi = \pm \phi_s(T)$. Within this curve, a uniform initial state $\bar\phi$ is locally unstable to perturbations. On the other hand, between the spinodal and the binodal ($\phi_s\le|\phi|\le\phi_b$) the uniform state is metastable; to get started, phase separation requires nucleation of a large enough droplet. For a further discussion on the thermodynamics of binary fluid mixtures, see \cite{chaikin,onuki}. 

\section{Mean field theory}
The simplest approach to the binary fluid is to postulate the following Landau-Ginzburg free energy functional
\begin{equation}
F[\phi] = \int dV\left(\frac{a}{2}\phi^2+\frac{b}{4}\phi^4 +\frac{\kappa}{2}(\nabla\phi)^2\right)
\label{functional}
\end{equation}
where $b$ and $\kappa$ are positive.
Technically this is an expansion about a critical temperature $T_C$, at which the parameter $a$ changes sign (positive above $T_C$, negative below). More importantly, apart from certain details at low temperature (where it does not predict correctly the exponential approach of $\phi_b\to 1$), this functional can describe not only the physics of the phase diagram in Fig.\ref{one}, but also, in schematic form, the physics of interfacial tension. These issues can be tackled without further approximation by addressing (\ref{functional}) using field theory methods (the renormalization group being essential to understand the behaviour around $T_C$) but for our purposes, mean-field theory is sufficient. The mean-field theory is found simply by minimizing $F$. 

Two comments are in order. First, it would be possible to add a linear term $\int \phi dV$ to $F$. However, this equates to $(\bar{n}_A-\bar{n}_B)/(\bar{n}_A+\bar{n}_B) = \bar\phi V$ which is simply a constant governed by the global composition. Since it cannot vary, this term in $F$ is physically irrelevant. Second, for a general fluid mixture one can expect a cubic term $\int (c\phi^3/3) dV$. This term creates an asymmetric phase diagram, and is clearly important in fitting the model to real fluid pairs for which some asymmetry is always seen. However, for the physics discussed in these lectures, the cubic term adds lots of algebra and not much physics, so we omit it.

\begin{figure}[hbtp]
	\centering
	\includegraphics[width=11.truecm]{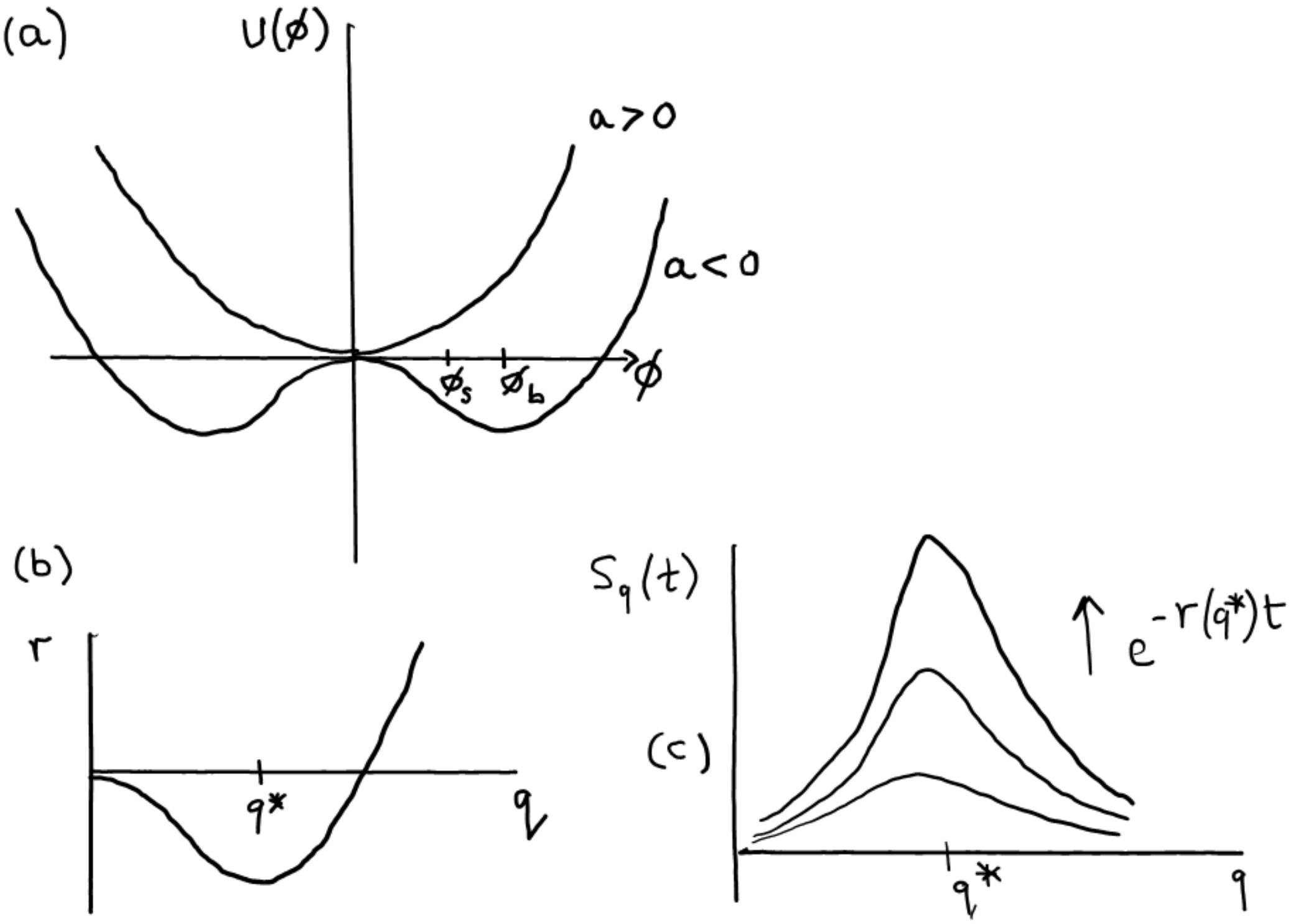}
	\caption[]{(a) Local part of free energy density $U(\phi)$ for $a>0$ and $a<0$. (b) The decay rate $r(q)$ in the spinodal regime. (c) Resulting growth of a peak in the equal time density correlator $S_q(t)$.}
	\label{two}
\end{figure}

The mean-field approach to (\ref{functional}) first considers states of uniform $\phi(\r) = \bar\phi$. For such states
\begin{equation}
\frac{F}{V} = \frac{a}{2}\bar\phi^2+\frac{b}{4}\bar\phi^4 \equiv U(\bar\phi) \label{uniform}
\end{equation} 
This has a single minimum at $\bar\phi = 0$ for $a>0$, with positive curvature everywhere (Fig.\ref{two}). The latter means that whatever $\bar\phi$ is chosen, one cannot lower the free energy by introducing a phase separation. On the other hand, for $a<0$, $F/V$ has negative curvature at the origin (and indeed everywhere between the spinodals $\phi_s = (-a/3b)^{1/2}$). Moreover it has two symmetric minima at $\bar\phi = \pm\phi_b$ with $\phi_b = (-a/b)^{1/2}$. For $|\bar\phi|<\phi_b$, $F/V$ is minimized by demixing the uniform state at $\bar\phi$ into two coexisting states at $\phi = \pm\phi_b$. (A small price must be paid to create an interface between these, but in the thermodynamic limit it is always worth paying.) The phase volumes of the bulk coexisting states is given by (\ref{phasevol}). 

\section{Interfacial tension}
These two bulk phases will organize themselves to minimize their mutual surface area; in most geometries, this requires the interface to be flat. To calculate its interfacial tension, we need to know the interfacial profile. We take a flat interface with its normal along the $x$ direction so that $\phi(\r) = \phi(x)$. The boundary conditions are that $\phi(x)$ approaches $\pm\phi_b$ at $x=\pm\infty$. To find the profile, we minimize $F[\phi]-\lambda\int\phi dV$ with these boundary conditions. (The $\lambda$ term ensures that the global composition remains fixed during the minimization.) 
The resulting condition
\begin{equation}
\frac{\delta}{\delta\phi}\left[F-\lambda\int \phi dV\right] = 0
\end{equation}
shows that the chemical potential $\mu \equiv \delta F/\delta\phi$ is equal to $\lambda$ and hence independent of position. Since for our symmetric choice of $F[\phi]$ we have $\mu = dU/d\phi = 0$ in a uniform bulk phase at density $\pm\phi_b$, it follows that $\lambda = 0$ and
\begin{equation}
\mu(x) = a\phi + b\phi^3 -\kappa\nabla^2\phi  = 0 \label{chempot}
\end{equation}
This expression for $\mu$ follows from the variational calculus \cite{VC}, which evaluates the relevant functional derivatives as
\begin{eqnarray}
\frac{\delta}{\delta\phi(\r)}\int\phi^ndV&=& n\phi(\r)^{n-1}\\
\frac{\delta}{\delta\phi(\r)}\int(\nabla\phi)^2dV &=& -2\nabla^2\phi(\r)
\end{eqnarray}

It is then a mathematical exercise \cite{chaikin} to show that, with the boundary conditions already given, the solution of (\ref{chempot}) is
\begin{equation}
\phi_0(x) = \pm\phi_b\tanh\left(\frac{x-x_0}{\xi_0}\right) \label{profile}
\end{equation}
Here $\xi_0 = (-\kappa/2a)^{1/2}$ is an interfacial width parameter, $x_0$ marks the midpoint of the interface, and the overall sign choice depends on whether the A-rich or B-rich phase occupies the region at large positive $x$. 

Likewise one may show that the interfacial tension, defined as the excess free energy per unit area of a flat interface over the bulk phases, obeys \cite{chaikin}
\begin{equation}
\gamma_0(x) = \int_{-\infty}^{\infty} \left(U(\phi_0)
+\frac{\kappa}{2}\phi_0^2-U(\phi_b)\right)dx = \left(\frac{-8\kappa a^3}{9b^2}\right)^{1/3}\label{tension}
\end{equation}
where $\phi_0$ stands for $\phi_0(x)$ obeying (\ref{profile}). The first two terms in the integrand give the full interfacial free energy evaluated for the equilibrium profile and the third subtracts off the bulk contributions. The interfacial profile is fixed by a trade-off between the penalty for sharp gradients (set by $\kappa$) and the purely local free energy terms which, on their own, would be minimized by a spatial composition that jumps discontinuously from one bulk value to the other. All of the free energy parameters are ultimately controlled by molecular physics, but this dependence is system specific and we do not discuss it here.

\section{Stress tensor}
If the interfacial profile departs from the equilibrium one, a thermodynamic stress $\bsigma$ will act on the fluid. An important example is when the interface is not flat but curved; under these conditions $\mu$ cannot be zero everywhere.
For use in the NSE we require not the stress tensor directly but the thermodynamic force density ${\bf f} = \nabla.\bsigma$ which, as stated in (\ref{body}) can also be expressed as $-\phi\nabla\mu$. 

One derivation of this result is to consider moving a small blob of fluid of volume $\Delta V$ and composition $\phi$, from one part of the system to another. Denoting the chemical potential at its original position by $\mu$ and that at its final position by $\mu+\delta\mu$, the free energy change $\delta F$ is (arguably) $\phi\delta\mu \Delta V$.
If this happens within a region where $\mu$ varies slowly, so that $\delta\mu = \nabla\mu.\delta\r$, we have
\begin{equation}
\delta F = \Delta V\,(\phi \nabla\mu).\delta\r \label{quick}
\end{equation}
This is the work done by an external agent who must therefore apply a force density $\phi\nabla\mu$. Thus the internal force density exerted by the surroundings on the blob is
\begin{equation}
{\bf f} = -\phi\nabla\mu \label{body2}
\end{equation}
which is the required result. 

The above argument is seductive but incomplete: we are working at fixed fluid density, so some unspecified other blob of fluid must move in the opposite direction from the one we have transported. A more correct, but much longer, derivation is to consider an incompressible deformation of the binary fluid mixture, in which line elements transform as $dr_i\to dr_i+\epsilon_{ij}dr_j$, with $\epsilon_{ij}$ the strain tensor. This deformation moves fluid elements about but cannot change their local composition: if a material point moves from $\r$ to $\r'$ then $\phi(\r')$ after deformation equals $\phi(\r)$ before. This information is sufficient to allow the free energy functionals $F[\phi]$ before and after the deformation to be compared; to leading order in $\epsilon_{ij}$ one has
\begin{equation}
\Delta F = \frac{1}{2}\sigma_{ij}\epsilon_{ij}
\end{equation}
The resulting stress tensor is found (eventually) to be \cite{reichl}
\begin{equation}
\sigma_{ij} = \left[-\frac{a}{2}\phi^2-\frac{3b}{4}\phi^4 +\kappa\phi\nabla^2\phi + \frac{1}{2}\kappa(\nabla\phi)^2\right]\delta_{ij} -\kappa(\partial_i\phi)(\partial_j\phi)
\end{equation}
from which (\ref{body2}) can be confirmed (this is left as an exercise).

\section{Equation of motion for composition; Model H}\label{ModH}
Having specified how to compute the force term in the NSE for a binary fluid from the composition field $\phi(\r)$, we now need an equation of motion for this quantity itself. This takes the form
\begin{equation}
\dot\phi +\v.\nabla\phi = - \nabla.{\bf J} \label{current1}
\end{equation}
where the term in $\v$ represents advection by the fluid velocity, so the left-hand side is the co-moving time derivative of $\phi$. This derivative must be the divergence of a flux, because A and B particles are not created or destroyed and thus $\phi$ is a conserved field. The form chosen for the flux is
\begin{equation}
{\bf J}  = -M(\phi)\nabla\mu \label{current2}
\end{equation}
where $M(\phi)$ is a mobility that depends locally on composition. More generally one could have a nonlocal $M[\phi]$; but for our purposes it will generally be enough to consider $M(\phi)$ to be constant. This mobility factor describes, under conditions of fixed total particle density, how fast A and B molecules can move down their respective chemical potential gradients to relax the composition field. 

Combining (\ref{current1},\ref{current2}) with our earlier results for the chemical potential and the forced NSE, we arrive at a closed set of dynamical equations with which to describe the dynamics of phase separation in an incompressible, isothermal, binary fluid mixture: 
\begin{eqnarray}
\mu(\r) &=& a\phi(\r) + b\phi^3(\r) -\kappa\nabla^2\phi(\r)\\
\rho(\dot\v+\v.\nabla\v) &=& \eta\nabla^2\v-\nabla P - \phi\nabla\mu \label{HNSE}\\
\nabla.\v &=& 0 \label{incomp}\\
\dot\phi + \v.\nabla\phi &=& \nabla.(M\nabla\mu)\label{Hphi}
\end{eqnarray}
Collectively the above equations are known as `Model H' \cite{chaikin}. Note that the pressure field $P$ has the job of ensuring incompressibility so that (\ref{incomp}) is obeyed. In this constraint-enforcing role, $P$ takes on whatever value it is told to by the other terms in the equations and cannot therefore introduce any new scaling behaviour that was not already calculable from those terms. 

As we have derived them, the equations of Model H are of mean-field form: they are deterministic, and take no account of noise. Noise terms are however important in at least two situations. One is near the critical point (not addressed in these lectures) where thermal fluctuations play a dominant role in the statistics of $\phi$: the mean-field theory implicit in the noise-free treatment breaks down. Another case where noise matters is when one has suspended fluid droplets (and/or colloidal particles). These objects will move by Brownian motion, which arises from thermal momentum fluctuations. These are neglected by our noise-free NSE, so that droplets of one fluid in another cannot diffuse.

Fortunately the noise terms can easily be determined using the fluctuation dissipation theorem \cite{chaikin}. For the order parameter fluctuations we need to add to (\ref{Hphi}) a term $-\nabla.{\bf J}^n$ where the random current ${\bf J}^n$ (superscript $n$ for noise) has the following statistics:
\begin{equation}
\langle J^n_i(\r,t)J^n_j(\r',t')\rangle = 2k_BTM\delta_{ij}\delta(\r-\r')\delta(t-t')\label{HJnoise}
\end{equation}
Similarly, to include Brownian motion we need to add to 
(\ref{HNSE}) a term $-\nabla.\bsigma^n$ where $\bsigma^n$ is a fluctuating thermal stress whose statistics obey \cite{landau}
\begin{equation}
\langle \sigma^n_{ij}(\r,t)\sigma^n_{kl}(\r',t')\rangle = 2k_BT\eta\delta_{ik}\delta_{jl}\delta(\r-\r')\delta(t-t')\label{Hvnoise}
\end{equation}
With these terms added, Model H is transformed from a mean-field approximation to a complete description of the binary fluid whose free energy functional is $F[\phi]$. Accordingly, in the absence of flow and with noise terms included, Model H ultimately achieves in steady state the probability density ${\mathcal P}[\phi]\propto\exp\left(-\beta F[\phi]\right)$, with $\beta\equiv(k_BT)^{-1}$, as required by the Boltzmann distribution.

\chapter{Phase Separation Kinetics} \label{PSK}
Having assembled the conceptual and mathematical tools we need, let us now examine some of the dynamics of phase separation.

\section{Spinodal decomposition}
We start by considering the spinodal instability. Ignoring advection initially, we write (\ref{Hphi}) as 
\begin{eqnarray}
\dot\phi &=& \nabla.(M\nabla\mu)\\
&=& \nabla.\left(M\nabla\left[U'(\phi)-\kappa\nabla^2\phi\right]\right)\\
&=& \nabla.\left(M\left[U''(\phi)\nabla\phi-\kappa\nabla^2\phi\right]\right)
\end{eqnarray}
Here $U(\phi)$ is the local free energy density of a uniform state as defined in (\ref{uniform}), and primes denote differentiation of this function with respect to $\phi$. 

Next we linearize this equation about a uniform initial composition $\bar\phi$, and fourier transform, to give
\begin{equation}
\dot\phi_q = -M(\bar\phi)q^2\left[U''(\bar\phi)+\kappa q^2\right]\phi_q
= -r(q)\phi_q
\end{equation}
where the second equality defines a wave-vector dependent decay rate $r(q)$. For $U''(\bar\phi)>0$ this is positive for all $q$: all fourier modes decay and the initial state is stable. In contrast for $U''(\bar\phi)<0$, the system is unstable, with $r(q)$ negative at small and intermediate wavevectors. (Stability is restored at high enough $q$ by the $\kappa$ term.)
Differentiating the growth rate $-r(q)$ with respect to $q$ we can identify the fastest growing instability to be at $q^* = -U''(\bar\phi)/2\kappa$ (Fig.\ref{two}). 

Even neglecting the noise terms in the dynamics (\ref{HJnoise},\ref{Hvnoise}), the initial condition can be assumed to have some fluctuations. Those whose wavenumber lies near $q^*$ grow exponentially faster than the rest, so that the time dependent composition correlator $S_q(t) =\langle\phi_q(t)\phi_{-q}(t)\rangle$ soon develops a peak of height scaling as $\exp[|r(q^*)|t]$ around $q^*$ (Fig.\ref{two}). Hence during this `early stage' of spinodal decomposition a local domain morphology is created by compositional diffusion (inter-diffusion of A and B) with a well defined initial length scale set by $\pi/q^*$. The amplitude of these compositional fluctuations grows until local values approach $\pm\phi_b$, the composition at which A-rich and B-rich phases can coexist. There soon develops a domain pattern, still initially with the same length scale, consisting locally of these phases, separated by sharp interfaces whose local profiles resemble (\ref{profile}).

\section{Laplace pressure of curved interfaces}
As mentioned previously, however, unless these interfaces are perfectly flat, they will exert forces on the fluid via the $-\phi\nabla\mu$ term in (\ref{HNSE}), in response to which the fluid will be set in motion. The physics of this term, for interfaces that are locally equilibrated but not flat, is that of Laplace pressure. 

To remind ourselves of this physics, consider a spherical droplet of one fluid in another with radius $R$ and interfacial tension $\gamma$. Let the pressure inside the droplet be greater than that outside by an amount $\Delta P$. The total force on the upper half of the droplet exerted by the bottom half is then
\begin{equation}
\pi R^2\Delta P -2\pi \gamma R = 0
\end{equation}
which must vanish if the droplet is not moving. The first term comes from the vertical component of the extra pressure acting across the equatorial disc, and the second is the tension acting across its perimeter. Hence $\Delta P = 2\gamma/R$; equilibrium requires the internal pressure to be higher as a result of curvature and interfacial tension. More generally one has a Laplace pressure
\begin{equation}
\Pi = \gamma\left(\frac{1}{R_1}+\frac{1}{R_2}\right)
\end{equation}
where $R_1$ and $R_2$ (see Section \ref{tensionless}) are the principle radii of curvature of the interface.

\section{What happens next?}
The next stage of the phase separation kinetics depends crucially on the topology of the newly formed fluid domains. This is controlled mainly by the phase volumes $\Phi_{A,B}$ of the A-rich and B-rich phases. Roughly speaking, if $0.3\le\Phi_A\le 0.7$ the domain structure will be bicontinuous: one can trace a path through the A-rich phase from one side of the sample to the other, and likewise for the B-rich phase. Outside this window, the structure instead has droplets of A in B ($\Phi_A<0.3$) or B in A ($\Phi_A>0.7$). The values $0.3$ and $0.7$ are rule-of-thumb figures only, with details depending on many other factors (including any asymmetry in viscosity \cite{onuki,coveney}) that we do not consider here. Note also that the window of bicontinuity shrinks to a single point in two dimensions, where the slightest asymmetry in phase volume and/or material properties will generally result in a droplet geometry in which only one phase is continuous. 

\section{Coalescence of droplet states}

If the post-spinodal structure is that of droplets, each relaxes rapidly to minimize its area at fixed volume resulting in a spherical shape. After this, Laplace pressures are locally in balance and although stresses are still present there is no net fluid motion in the absence of noise. Thermal noise however allows droplets to explore space by Brownian motion, and the resulting collisions cause the mean droplet radius $R$ to increase by coalescence. To estimate the rate of this process, we observe that for A droplets in B, the mean inter-droplet distance $L$ is of order $R\Phi_A^{-1/3}$ at small $\Phi_A$ or, more generally,
$L\sim Rf(\Phi_A)$. Each droplet will collide with another in a time $\Delta t$ of order $L^2/D$ where $D \simeq k_BT/\eta R$ is the diffusivity. Upon collision, two droplets of radius $R$ make a new one of radius $2^{1/3}R$ causing an increment $\Delta \ln R = (\ln 2)/3$. This gives
\begin{equation}
\frac{\Delta \ln R}{\Delta t} \propto \frac{k_BT}{\eta R^3}
\end{equation}
where the left hand side can now be approximated as $d\ln R/dt = \dot R/R$. By integration we then obtain the scaling law
\begin{equation}
R(t) \sim\left(\frac{k_BTt}{\eta}\right)^{1/3} \label{coalescence}
\end{equation}
This argument assumes that coalescence is diffusion-limited, and shows that in this case Brownian motion will cause indefinite growth of the mean droplet size, culminating in total phase separation. 

The assumption of diffusing spherical droplets is reliable at low phase volumes $\Phi_A$ of the dispersed phase, but when this is not small more complicated routes to coalescence, some involving droplet-scale or macroscopic fluid flow, are possible. One of these is so-called `coalescence-induced coalescence' where the shape relaxation post-collision of a pair of droplets creates enough flow to cause another coalescence nearby \cite{CIC}. This gives a new scaling ($R\sim \gamma t/\eta$) which coincides with one of the regimes described later for the coarsening of bicontinuous structures ((\ref{VH}) below) and indeed stems from the same balance of forces as will be discussed for that case.

In many droplet emulsions it is possible to inhibit the coalescence step, so that this route to phase separation is effectively blocked. For instance, adding charged surfactants can stabilize oil droplets in water against coalescence by creating a coulombic barrier opposing the close approach of droplet surfaces. Steric interactions between surfactant tails can likewise stabilize water-in-oil emulsions. If the rupture of a thin film of the continuous phase between droplets has a high enough nucleation barrier, coalescence rates can often be reduced to a manageable or indeed negligible level \cite{bibette}.

\section{Ostwald ripening}
Sadly however, switching off coalescence is not enough to prevent macroscopic phase separation of droplet emulsions. This is because of a process called Ostwald ripening, in which material is transported from small droplets to large ones by molecular diffusion across the intervening continuous phase. 

This process is most easily considered when $\phi_b$ is close to unity so that A droplets consist of nearly pure A. There is nonetheless a small equilibrium concentration $c_{eq}$ of A molecules in the nearly pure B phase: this is what allows diffusion of A between droplets. 

The driving force for Ostwald ripening is the Laplace pressure difference between small and large droplets. In a droplet of radius $R$ the Laplace pressure $\Pi = 2\gamma/R$ causes the chemical potential $\mu_A$ of A molecules within it to be raised by $\Delta \mu_A(R) = \Pi v_A$ where $v_A$ is a molecular volume. Because the interface is in local equilibrium, the chemical potential just {\em outside} a droplet of radius $R$ is also raised by this amount. Treating the dilute solution of A in B exterior to the droplet as an ideal mixture, we then have a concentration field $c_A(r)$ at radius $r$ from the droplet centre that obeys 
\begin{equation}
c_A(R^+) = c_{eq}\left(1+\frac{2\gamma v_A}{k_BT R}\right) \label{Ost1}
\end{equation} 
This concentration is high outside small droplets and lower outsider large ones; the resulting gradient causes a diffusive flux from smaller to larger droplets. 

To address this at mean field level we consider just one droplet, of radius $R$, obeying
(\ref{Ost1}) and take a boundary condition at infinity
\begin{equation}
c_A(r\to\infty) = \bar c(t) = c_{eq}(1+\epsilon(t)) \label{Ost2}
\end{equation}
Here the $\epsilon(t)$ term denotes the fact that the system globally has not yet reached equilibrium: hence there is a mean supersaturation of A in B resulting from the fact that distant droplets are themselves of finite radius. (Note that $\epsilon(t)$ will tend to zero if and when the mean droplet size tends to infinity.)

\begin{figure}[hbtp]
	\centering
	\includegraphics[width=11.truecm]{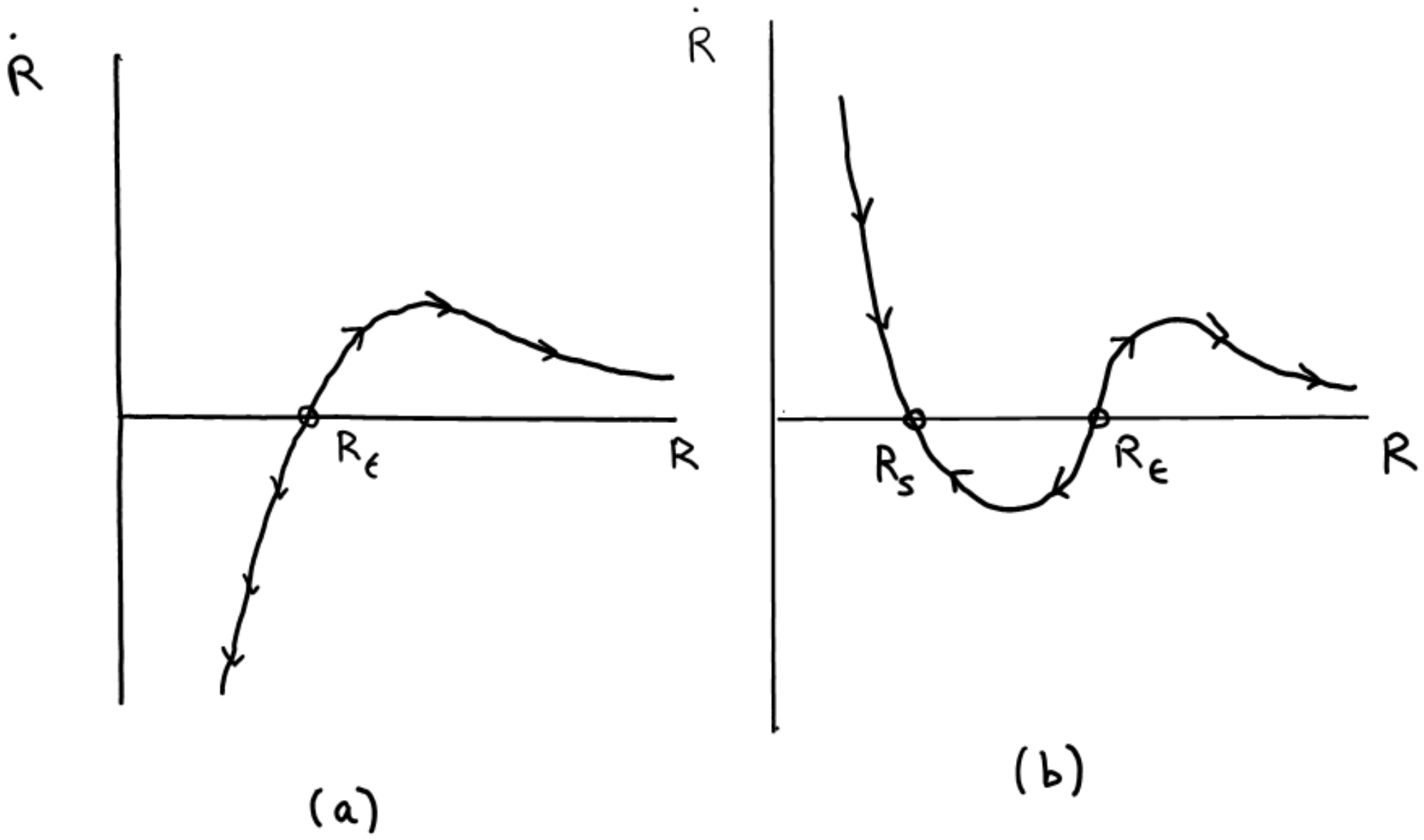}
	\caption[]{Growth rate of droplet as a function of size during the Ostwald process (a) without and (b) with trapped species.}
	\label{three}
\end{figure}

It is a simple exercise then to solve the quasi-steady diffusion equation $D_A\nabla^2c_A = 0$ with the boundary conditions (\ref{Ost1},\ref{Ost2}) to find
\begin{equation}
c_A(r) =  \bar c(t) + R\left(\frac{c(R^+)-\bar c}{r}\right)\label{Ost3}
\end{equation}
and from this to derive the flux of A molecules onto (or off of) the droplet surface. The result is an equation for the droplet size
\begin{equation}
\dot R = \frac{v_AD_Ac_{eq}}{R}\left(\epsilon-\frac{2\gamma v_A}{k_BTR}\right) \label{LSgrowth}
\end{equation}
The function $\dot R(R)$ is shown in Fig.\ref{three}. This exhibits an unstable fixed point at 
\begin{equation}
R = R_\epsilon(t)\equiv \frac{2\gamma v_A}{\epsilon k_B T}
\end{equation}
Droplets bigger than this grow, those smaller, shrink. 

We can now find the scaling of the typical droplet size by assuming this to be comparable (but not exactly equal) to $R_\epsilon$:
\begin{equation}
\dot R \simeq \frac{v_AD_Ac_{eq}}{R}\frac{\gamma v_A}{k_BTR}
\end{equation}
giving the scaling law \cite{bibette}
\begin{equation}
R(t) \simeq \left(\frac{v^2_AD_Ac_{eq}t\gamma}{k_BT}\right)^{1/3}
\sim t^{1/3}\label{LSlaw}\end{equation}
From this it follows that the global supersaturation varies as $\epsilon(t)\sim t^{-1/3}$. A more complete theory, due to Lifshitz and Slyozov, not only confirms these scalings but gives detailed information on the droplet size distribution \cite{LSW}.
Note that (\ref{LSlaw}) has similar time dependence to (\ref{coalescence}) for coalescence; this stems from the fact that both mechanisms are ultimately diffusive. However, the nature of the diffusing species (droplet in one case, molecule in the other) is quite different, resulting in prefactors that involve unrelated material properties for the two mechanisms.

\section{Preventing Ostwald ripening}\label{ts}
We see from (\ref{LSlaw}) that the Ostwald process can be slowed by reducing the interfacial tension $\gamma$. This is discussed later in relation to surfactants, but unless the tension is reduced effectively to zero this will only slow things down by a moderate factor. One might also reduce the solubility $c_{eq}$ of A in B, but generally the aim is to make emulsions of particular fluids, so replacing A with a less soluble species is often an impractical suggestion. On the other hand, a closely related approach is to include within the A phase a modest concentration of a species that is effectively insoluble in B. This might be a polymer or, if A is water and B oil, a simple salt. The idea is that the trapped species in the A droplets creates an osmotic pressure which rises as $R$ falls, hence opposing the Laplace pressure. Treating the trapped species as an ideal solution in A,  (\ref{LSgrowth}) is replaced by \cite{webster}
\begin{equation}
\dot R = \frac{v_AD_Ac_{eq}}{R}\left(\epsilon-\frac{2\gamma v_A}{k_BTR}+\frac{\zeta v_A}{(4\pi/3)R^3}\right) \label{LSgrowth2}
\end{equation}
where the last term is $\Pi_O v_A/k_BT$ with $\Pi_O$ the osmotic pressure of $\zeta$ trapped particles within a droplet of radius $R$. Figure \ref{three} shows the new structure of the $\dot R(R)$ equation. 

There is now a stable fixed point at a size determined by the trapped species
\begin{equation}
R_\zeta = \left(\frac{3\zeta k_BT}{8\pi\gamma}\right)^{1/2}
=\left(\frac{c_T R_0^3 k_BT}{2\gamma}\right)^{1/2}
\end{equation}
where $R_0$ is the initial droplet size and $c_T$ the initial concentration of trapped species in the A phase.
Droplets that have shrunk to a size $R_\zeta(R_0)$ can coexist with a bulk phase of $A$ without shrinking further: the Laplace pressure is balance by $\Pi_O$. Indeed if the initial size obeys $R_0<R_\zeta(R_0)$ (treating the droplets as monodisperse for simplicity) the Ostwald process is switched off entirely. By this route one can thus make robust `mini-emulsions' \cite{mini} or `nano-emulsions' \cite{nano}, which will not undergo coarsening by the Ostwald process. However these are still metastable: so long as the tension $\gamma$ is positive, the free energy can always be reduced by coalescing droplets to reduce the interfacial area. 

\section{Coarsening of bicontinuous states}
As mentioned previously, for $0.3\le\Phi_A\le 0.7$ (roughly speaking) the domains of A-rich and B-rich coexisting fluids remain bicontinuous. This allows coarsening by a process faster than either coalescence or Ostwald ripening, in which the Laplace pressure gradients pump fluid from one place to another. The driving force is interfacial tension and at any time there is a characteristic domain length scale $L(t)$, much larger than the interfacial width, which we shall assume to be the only relevant length in the problem, so that $\nabla\sim 1/L$. Moreover the characteristic magnitude of the Laplace pressure is $\Pi\sim\gamma/L(t)$. This fixes the scale of the forcing term in the Navier Stokes equation (\ref{HNSE}) as $-\phi\nabla\mu\sim\nabla\Pi\sim \gamma/L^2$. The fluid velocity is of order $\dot L$ so the viscous term scales as $\eta \dot L/L^2$. The inertial terms are $\rho\dot\v \sim \rho \ddot L$ and $\rho \v.\nabla\v \sim \rho (\dot L)^2/L$. The $\nabla P$ term, which ensures incompressibility, is slave to the other terms (see Section \ref{ModH}).

An important property of (\ref{HNSE}), once sharp interfaces are present so that $\phi\nabla\mu\sim \gamma/L^2$, is that it contains only three parameters, $\rho, \gamma, \eta$. From these three quantities one can make only one length, $L_0 = \eta^2/\rho\gamma$, and one time $t_0 = \eta^3/\rho\gamma^2$. This means that the domain scale $L(t)$ must obey \cite{siggia,furukawa,kendon}
\begin{equation}
\frac{L(t)}{L_0} = f\left(\frac{t}{t_0}\right)
\end{equation} 
where, for given phase volumes, $f(x)$ is a function common to all symmetric binary fluid pairs. Note that noise is excluded, so that one expects this scaling to fail in droplet regimes where diffusion is important. (Noise may also be important in 2D \cite{jy}.) It can also fail in bicontinuous states at relatively early times and very high viscosities when an Ostwald-like process can dominate over fluid flow, giving $t^{1/3}$ scaling as in (\ref{LSlaw}). 

Excluding that regime, in bicontinuous states we therefore expect different behaviour according to whether $L/L_0$ is large or small. For $L/L_0$ small it is simple to confirm that the inertial terms in (\ref{HNSE}) are negligible. (Note also that, in any regime where $f(x)$ is a power law, the two inertial terms have the same scaling.) The primary balance in the NSE is then $\eta \dot L/L^2 \sim \gamma/L^2$ resulting in the scaling law $L(t) \sim \gamma t/\eta$ so that
\begin{equation}
f(x) \propto x\;\;;\;\; x\ll x^* \label{VH}
\end{equation} 
This is called the viscous hydrodynamic or VH regime \cite{siggia}. In contrast, at large $x$ the primary balance in the NSE is between the interfacial and inertial terms. It is a simple exercise then to show that $L(t) \sim (\gamma/\rho)^{1/3}t^{2/3}$ so that 
\begin{equation}
f(x) \propto x^{2/3}\;\;;\;\; x\gg x^* \label{IH}
\end{equation} 
This is called the inertial hydrodynamic (IH) regime \cite{furukawa}. 
In (\ref{VH},\ref{IH}) we have introduced a crossover value $x = x^*$ between the VH and IH regimes. In practice this crossover is very broad, and the crossover value rather high: $x^*\simeq 10^4$. The high crossover point is less surprising if one calculates a domain-scale Reynolds number
\begin{equation}
\mbox{\rm Re} = \frac{\rho L \dot L}{\eta} = f(x)\frac{df}{dx}
\label{Re}\end{equation} 
The crossover value of Re then turns out to be of order 10 \cite{kendon}, and the largeness of $x^*$ is found to stem from a small constant of proportionality in (\ref{VH}). It means that in practice a clean observation of the IH regime has only been achieved in computer simulation: in terrestrial laboratory experiments the domains are by then so large that the slightest density difference between A and B causes gravitational terms to dominate.

\section{Shearing binary fluids}
The coarsening of bicontinuous demixed states described above leads inexorably to complete phase separation; in practice this is something that, as explained in the introduction, we often wish to avoid. In a processing context, it is sometimes enough to temporarily maintain a well-mixed, emulsified state merely by stirring the system. Though industrial stirring is complicated, for our purposes it is enough to consider the effects of a simple shear flow.

Consider such a flow with macroscopic velocity along $x$, and its gradient along $y$; $z$ is then the neutral (or vorticity) direction. In simulations one can use boundary conditions with one static wall at $y=0$, another sliding one at $y = \Lambda$ and periodic BCs in $x,z$ -- or in practice there are ways to introduce periodic BCs also in $y$. Nonetheless, the system size in that direction, $\Lambda$ is important in what follows. The top plate moves with speed $\Lambda/t_s$ where $1/t_s$ (usually denoted $\dot\gamma$) is the shear rate. 

The question we ask is whether nonequilibrium steady states now exist for which the fluid domains have finite length scales $L_{x,y,z}$ in all three directions. The simplest hypothesis is that these lengths, if they exist, all have similar scaling: $L_{x,y,z}\sim L$. Moreover, given the preceding discussion of terms in the NSE we expect in steady state that $L/L_0$ is now a function, not of $t/t_0$ but of $t_s/t_0$. That is, the previous dependence on time is no longer present (because we assume a steady state exists) but is replaced by a dependence on the inverse shear rate. The functional form of this dependence could in principle be anything at all, but the simplest scaling ansatz is that the system coarsens as usual until $t\sim t_s$, whereupon the shearing takes over and $L$ stops increasing. If so, 
\begin{equation}
\frac{L}{L_0} \simeq f\left(\frac{t_s}{t_0}\right)
\end{equation}
where $f(x)$ is {\em the same} function as introduced previously, for which (\ref{VH},\ref{IH}) hold. If so, for $t_s/t_0 \ll x^*$ we have $L/L_0 \sim t_s/t_0$ and for $t_s/t_0 \gg x^*$ we have $L/L_0\sim (t_s/t_0)^{2/3}$. 
The Reynolds number obeys Re$\sim f(x)df/dx$ as in (\ref{Re}). In contrast to what happens for any problem involving shear flow around objects of fixed geometry, Re is now {\em small} when the shear rate is {\em large} and vice versa. This is because at small shear rates, very large domains are formed.

The above picture is the simplest possible \cite{doi}. We note only two of several possible complications. First, in principle $L_{x,y,z}$ could all have different scalings. The resulting anisotropies could spoil any clear separation of the VH and IH regimes; with three-way force balance in the NSE, there is no reason to expect clean power laws for any of these quantities. Secondly, at high shear rates the system-size Reynolds number Re$_\Lambda \sim \rho \Lambda^2/\eta t_s$ becomes large. The presence of a complex microstructure could promote any transition to conventional fluid turbulence expected in this regime. 

In practice, experimental tests of the above predictions are patchy (see \cite{onuki} and references cited in \cite{strat}). Gravity complicates matters, as does viscosity asymmetry (unavoidable in practice) between phases. However, relatively clean tests are possible via computer simulation \cite{strat,stansell}. In 2D one finds apparent scaling laws $L_x/L_0\sim (t_s/t_0)^{2/3}$ and $L_y/L_0\sim (t_s/t_0)^{3/4}$ over a fairly wide range of length and timescales, most of which are however in the crossover region around $x^*$ \cite{stansell}. The fitted exponents change slightly if instead of the flow and gradient direction one uses the principal axes of the distorted density patterns, but still do not coincide. In 3D, where the simulations require very large computations, the IH (2/3 power) scaling has been observed within numerical error for all three length scales within a range of accessible domain-scale Reynolds numbers Re$_L$
between 200 and 2000 \cite{strat}. These measurements are however limited by the onset of a macroscopic instability to turbulent mixing at Re$_\Lambda\simeq 20,000$. 

\begin{figure}[hbtp]
	\centering
	\includegraphics[width=10.truecm]{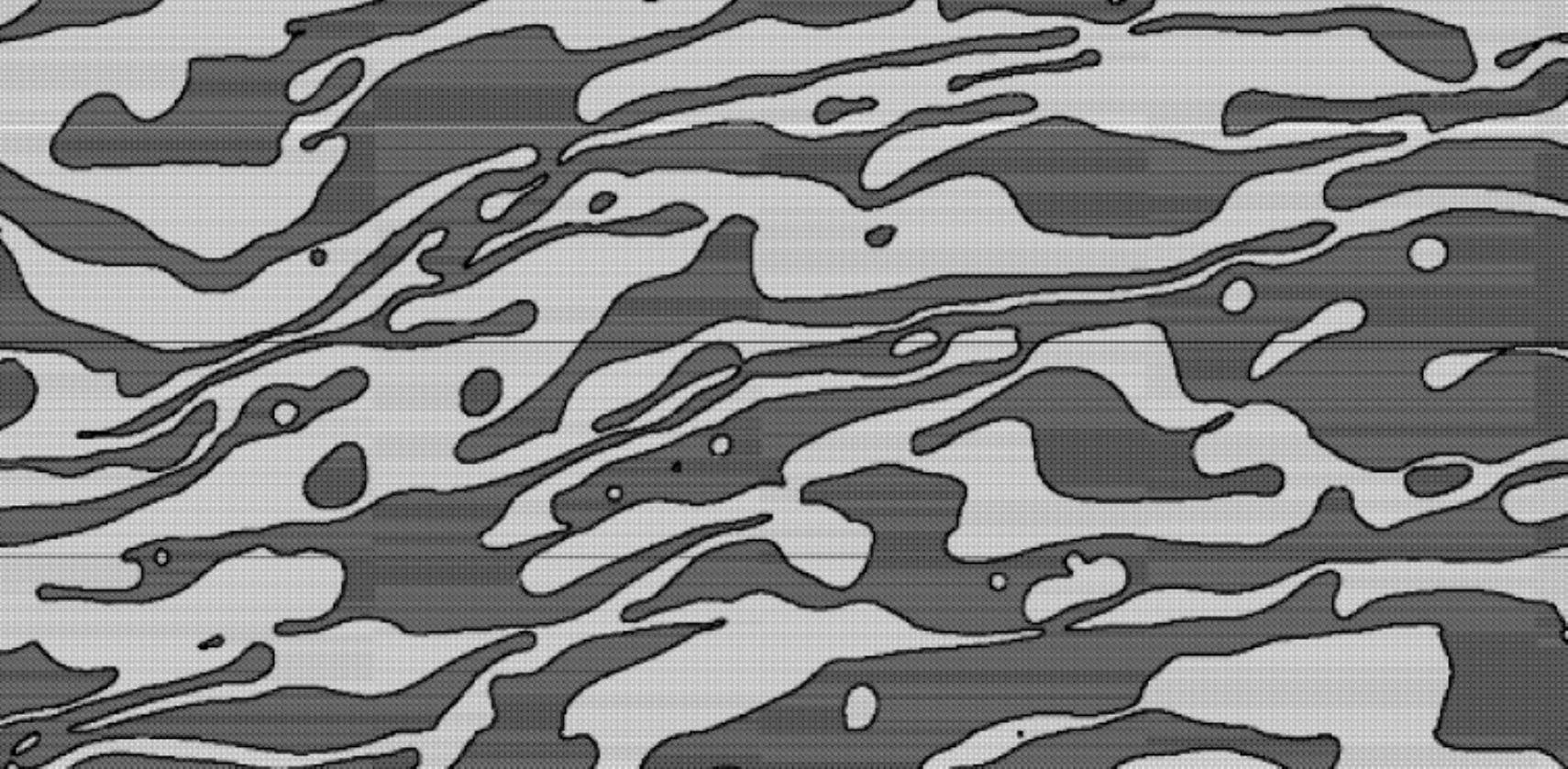}
	\caption[]{Image of binary fluid domains in a nonequilibrium steady state created by shearing. This is a 2D simulation, but a slice through a 3D run looks very similar. See \cite{strat,stansell}.}
	\label{four}
\end{figure}

A snapshot of the highly distorted domain structure seen in computer simulations of sheared binary fluids (within the laminar flow regime) is shown in Fig. \ref{four}. 

\chapter{Stabilizing Emulsions Thermodynamically}
Surfactants, comprising small amphiphilic models with a polar head group and an apolar tail, are well known to reduce the interfacial tension between coexisting phases of oil and water (as well as other pairs of apolar and polar fluids). Moreover they generally have fast exchange kinetics between the interface and at least one bulk phase in which they are soluble; this means that the interface remains locally in equilibrium, even if at larger scales the emulsion is only metastable. Use of surfactants can thus be viewed as a thermodynamic route to the stabilization of interfacial structures.

\section{Interfacial tension in the presence of surfactant}
\begin{figure}[hbtp]
	\centering
	\includegraphics[width=11.truecm]{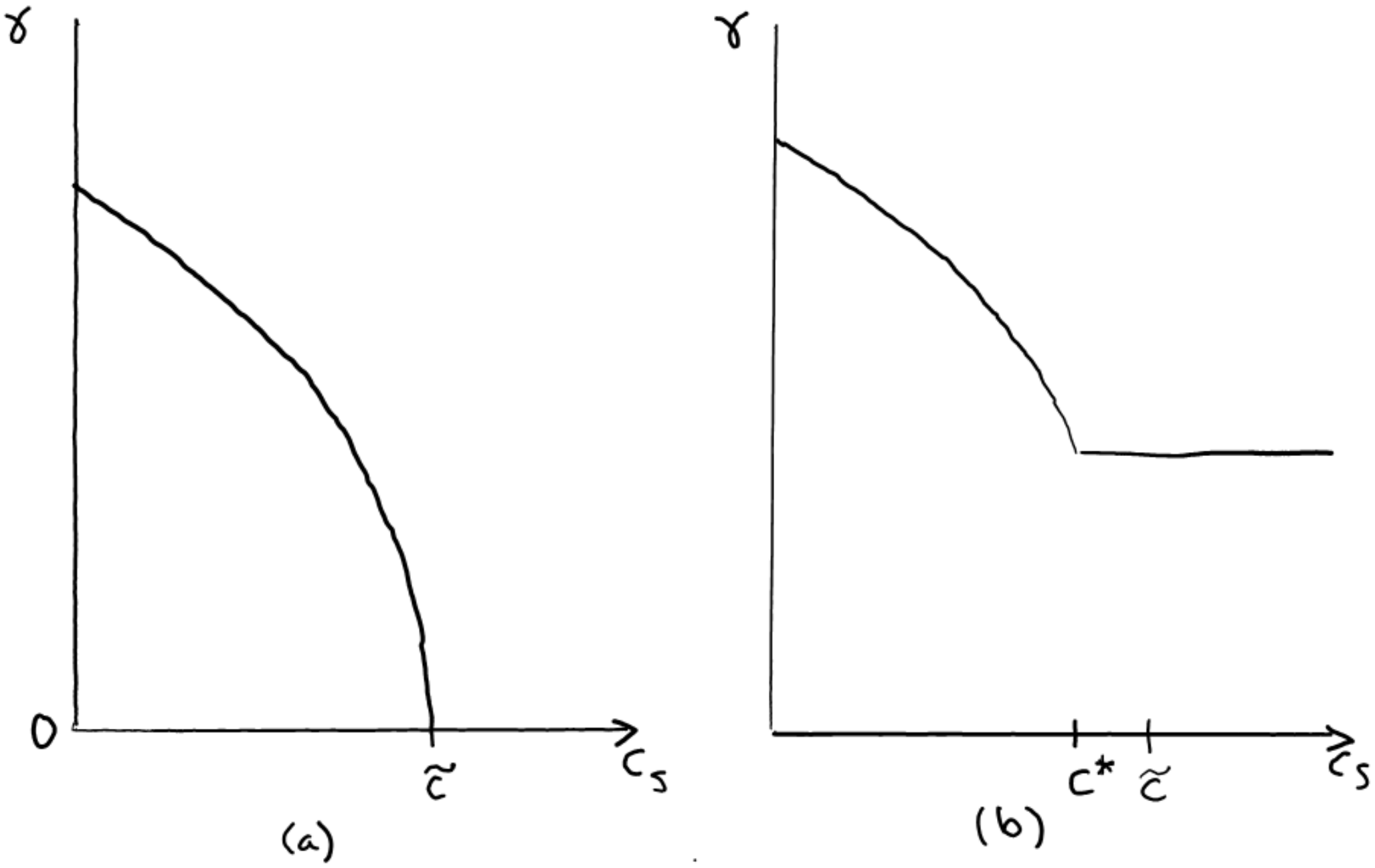}
	\caption[]{Interfacial tension as a function of surfactant concentration: (a) for an ideal surfactant solution, and the case where $c^*>\tilde c$; (b) for a typical surfactant where $c^*<\tilde c$, so that micellization pre-empts the vanishing of the interfacial tension $\gamma$. }
	\label{five}
\end{figure}

The basic effect can be illustrated by first considering an ideal solution of surfacant molecules each carrying a unit polarization vector $\p_i$. (This simply denotes orientation; a coulombic dipole is not required.) A local coarse graining creates a smooth field $\p(\r) = \langle\p_i\rangle_{meso}$; because the solution is ideal, it is easy to show that the variance $\chi = \langle|\p|^2\rangle$ of the fluctuating $\p$ field is proportional to the global concentration $c_s$ of the surfactant molecules. Therefore for noninteracting (ideal solution) surfactants one can write down a free energy to describe these fluctuations as
\begin{equation}
F_{s} = \int \left(\frac{|\p(\r)|^2}{2\chi}\right)dV\label{chidef}
\end{equation}
where (in terms of a concentration-independent parameter $\tilde\chi$)
\begin{equation}
\chi(c_s) = \tilde \chi c_s\label{chic}
\end{equation}
 is an osmotic compressibility. We next add a coupling term to represent the reduction in free energy caused when a surfactant molecule resides at the A-B interface with its polarity suitably aligned along the composition gradient $\nabla\phi$:
\begin{equation}
F_{c} = \int \nu \p.\nabla\phi\,dV
\end{equation}
The remaining terms of the free energy are simply those of the Model H binary fluid mixture, from (\ref{functional}), so that we now have
\begin{equation}
F[\phi,\p] = \int dV\left(\frac{a}{2}\phi^2+\frac{b}{4}\phi^4 +\frac{\kappa}{2}(\nabla\phi)^2+\frac{1}{2\chi}|\p|^2 + \nu\p.\nabla\phi\right)
\label{functional2}
\end{equation}
It is a simple exercise to minimize this over $\p(\r)$ at fixed $\phi(\r)$, and a slightly more complicated one to explicitly integrate over the fluctuating $\p$ field by Gaussian integration
to obtain $e^{-\beta F[\phi]} = \int e^{-\beta F[\phi,\p]}{\mathcal D}\p$. The result of either calculation is to recover the original Model H free energy (\ref{functional}),  but with a renormalized square gradient coefficient
\begin{equation}
\kappa_r = \kappa - \nu^2\chi(c_s)\label{renorm}
\end{equation}
From this it follows that the interfacial tension varies with the concentration of an ideal surfactant as
\begin{equation}
\gamma(c_s) = \left(\frac{-8a^3(\kappa-\nu^2\tilde\chi c_s)}{9b^2}\right)^{1/2}
\end{equation}
This vanishes, with infinite slope, at a concentration $c_s = \tilde c = \kappa/\nu^2\tilde\chi$ (Fig.\ref{five}). 

The above calculation for an ideal solution of surfactant gives a simplified picture of how, on increasing surfactant concentration, the interfacial tension between two immiscible fluids A and B can vanish. (For a far more advanced discussion, see \cite{gom}.) Before considering what happens when this point is reached, we must discuss why, in practice, it very often is not reached. This is because surfactant solutions are nonideal, due to the phenomenon of micellization in which individual molecules become aggregated into micelles containing several tens of molecules. The effect of this process, as we shall see below, is to put a cap on the osmotic compressibility $\chi(c_s)$. This happens at a certain concentration $c^*$; only if this lies beyond $\tilde c$ as defined above does $\gamma$ come close to zero.

\section{Micellization}
As just explained, surfactant solutions are not ideal. First recall that in an ideal solution, the concentration and chemical potential are linked by
\begin{equation}
c_s = v_T^{-1}\exp[\beta\mu_s]
\label{ideal}
\end{equation}
where $v_T$ is a molecular volume. (This is the same expression as for an ideal gas where $v_T$ is instead the thermal de Broglie volume.) Next, note that the role of $\chi$ is to tell us the cost of pulling surfactant molecules out of solution to put them on the interface: this is indeed the meaning of (\ref{chidef}). It should be clear therefore that for a nonideal solution what matters is not $c_s$ but $\mu_s$: we therefore expect (\ref{chidef}) to be replaced by
\begin{equation}
\chi(c) = \tilde\chi v_T^{-1}\exp[\beta\mu_s]\label{chic2}
\end{equation}
The  remaining task is to understand how $\mu_s$ behaves in the presence of micellization. 

\begin{figure}[hbtp]
	\centering
	\includegraphics[width=12.truecm]{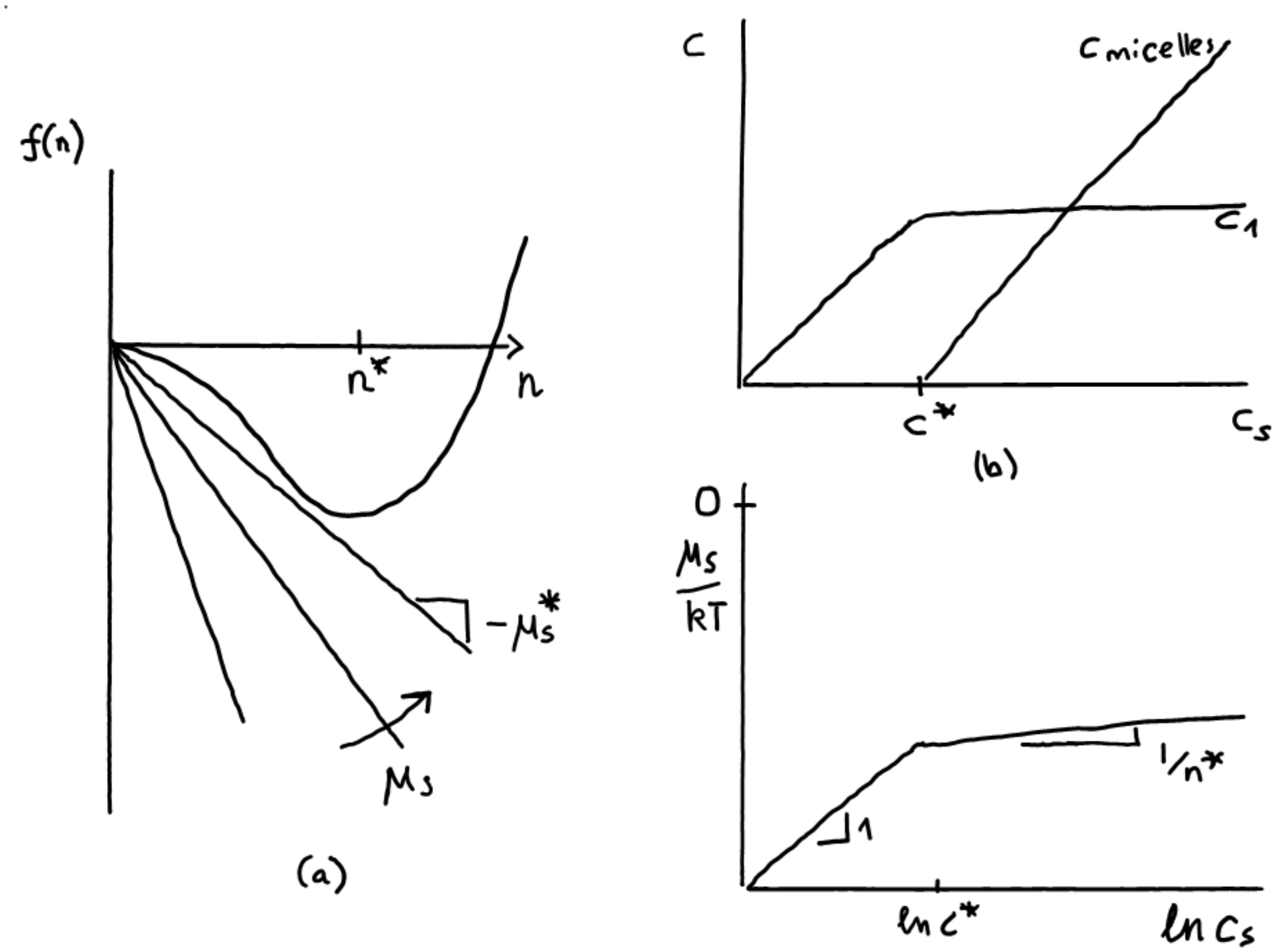}
	\caption[]{The physics of micellization. (a) Upper curve: 
the local free energy $f(n)$ as a function of aggregation number 
$n$ has negative curvature at small $n$ and positive at larger 
$n$, reflecting the existence of a preferred micellar size. The 
chemical potential $\mu_s$ is negative but rising with $c$: the 
three straight lines are examples of the function $\mu_s n$. The 
concentration $c(n)$ of $n$-mers is the exponential of the 
vertical separation between $f(n)$ and such a a straight line.
As $\mu_s$ increases, $c(n^*)$ is initially negligible but as 
$\mu_s\to\mu^*$ it abruptly overtakes the monomer concentration 
$c(1)$. (b) The resulting dependence on $c_s$ of both $c(1)$ and 
$c_{micelles} \simeq c_s-c(1)$. (c) The resulting dependence of 
$\mu_s$ 
on $c_s$, showing near-saturation at $c\ge c^*$. }
\label{six}

\end{figure}

The basic physics is shown in Fig. \ref{six}. As a function of the number of molecules $n$ in an aggregated cluster, the local free energy $f(n)$ of such a cluster first decreases slowly and then faster before increasing again beyond a characteristic size $n^*$. (We define $f(1) = 0$ as the baseline for local packing energy; by `local free energy' we meen a free energy that excludes the translational entropy of the micelle.) The initial downward curvature is because although two molecules can lower the energy of their hydrophobic tails somewhat by coming together, it is much more efficient to have a quorum of molecules so that the tails are entirely separated from water by a layer of heads. The upward curvature at $n\gg n^*$ is essentially because for larger $n$ there either has to be a hole in the centre of the micelle or it has to become aspherical. If the spherical packing is prefered, $f(n)$ is as sketched, but the effect on $\mu_s$ is no different if cylindrical micelles form instead. 

We account for micellization by treating the system as an ideal solution of aggregated objects ($n$-mers). According to the Gibbs-Boltzmann distribution the concentration of each sized aggregate then obeys
\begin{equation}
c(n) = v_T^{-1}\exp[-\beta(f(n)-n\mu_s)] \label{size}
\end{equation}
As shown in Fig. \ref{six}, for strongly negative chemical potential (which holds at low enough $c_s$) surfactant exists primarily as isolated molecules. As $\mu_s$ is raised, $c(1)$ increases monotonically. However, a point is eventually reached at which $f(n^*)-n^*\mu_s = \mu_s$ so that $c(n^*) = c(1)$.  By this point, since $n^*\gg 1$, almost all surfactant has formed micelles, and the total concentration $c_s$ is about $n^*$ times larger than $c(1)$. This situation is initially puzzling but has a simple explanation. For $\mu_s\ll\mu_s^*$ (where $\mu_s^*$ is defined roughly by the condition that the line $\mu_s^*n$ is tangent to the $f(n)$ curve) one has mostly unaggregated monomers with some exponentially rare micellar aggregates. As $\mu_s$ approaches $\mu_s^*$ (so that $\mu_s =\mu_s^*-\delta$, say), the monomer concentration $c(1)$ becomes essentially stuck at
\begin{equation}
c(1) = v_T^{-1}\exp[\beta\mu_s^*] \equiv c^*
\end{equation}
while any excess surfactant molecules beyond this ceiling form micelles (of size $n\simeq n^*$) instead. Because of the form of
(\ref{size}), the overall surfactant concentration can be raised by a large factor, of order $n^*$, for a very modest (order $k_BT$) increase in chemical potential. Thus, for practical purposes, it is a good approximation to say that $\mu_s$ hits a firm ceiling just below $\mu_s^*$, beyond which it effectively ceases to depend on $c_s$. Likewise the monomer concentration $c(1)$ saturates at $c^*$, which is commonly known as the {\em critical micelle concentration} or CMC. These behaviours are sketched in Fig.\ref{six}. In practice there is a rounded corner on $\mu_s$ and $c(1)$ rather than a singularity, and it is important to note that micellization represents merely a sharp crossover not a true phase transition. However, the curvature of these features diverges as $n^*\to\infty$.  In that limit the mathematics of micellization coincides precisely with that of Bose condensation for ideal quantum fluids, as addressed in standard statistical physics textbooks \cite{reichl}.

\subsection{Evolution of the interfacial tension}
According to the above arguments we now have two general classes of behaviour depending on whether the critical micelle concentration $c^*$ lies below or above $\tilde c$. Recall that this is the concentration of an ideal surfactant solution at which $\gamma$ would effectively vanish. In case 1, $c^*<\tilde c$, so that $\gamma(c_s)$ follows the ideal curve (Fig. \ref{five}) so long as $c_s<c^*$ but then abruptly stops decreasing as micellization intervenes: the chemical potential $\mu_s$ then saturates and no further decrease in tension is possible.  In case 2, $c^*>\tilde c$ so that $\gamma$ hits zero before micelles are formed. At this point, if water and oil are both present in bulk quantities, the system can minimize its free energy by creating a macroscopic amount of interface on which the surfactant can reside in comfort. When this happens $\mu_s$ again saturates: adding further surfactant simply creates more surface at fixed $\mu_s$. (Hence $\gamma$ can never actually become negative as the ideal solution calculation might suggest.) By this reasoning, micelles never form under `case 2' conditions.
 
The above represents an oversimplified, but useful, picture of the effects of surfactant on interfacial tension. In practice, contributions from curvature energy and entropy (considered below) may mean that interface forms spontaneously while $\gamma$ remains slightly positive. Indeed, we defined $\gamma$ as the tension of a flat interface; but the proliferation of interface occurs when the free energy cost vanishes for creating it not in that state, but the state of mimimum free energy (which might be crumpled or curved). Nonetheless, it is broadly correct to distinguish between case 1, where $\gamma$ is reduced considerably but remains or order its original value, and case 2 where $\gamma$ becomes effectively, if not actually, zero. (It is probably not correct to assume that micelles never form in case 2, however.) For a far fuller molecular discussion of how surfactants modify interfacial tension, see \cite{rosen}.

\section{Finite tension: metastable emulsions}
Because $c^*$ is generally small (typically $10^{-2}$ Mol l$^{-1}$, often far less), case 1 is generally the more common: the effect of surfactant is to reduce interfacial tension to half or a third of its previous value. Because $\gamma$ remains finite, the global minimum of free energy is always that of coexisting A-rich and B-rich phases, separated by a flat interface. The area of this interface is set by the container geometry (and, in practice, gravity) and scales as $V^{2/3}$ where $V$ is the sample volume. Interfacial physics thus contributes negligibly to equilibrium states in the thermodynamic limit. Nonetheless, one can create metastable emulsions, for instance by stirring. These are generally droplets (of A in B, say) but by drainage under gravity, for instance in a centrifuge, much of the continuous B phase can often be expelled to create a so-called biliquid foam \cite{bibette}. 

As the name suggests, biliquid foams are very similar in structure to foams made of air bubbles in a surfactant solution (soap froths). They comprise polyhedral droplets of A (say) separated by thin films of B, and the stability of the foam depends on a barrier to coalescence of A across these thin films. The role of surfactant is more to do with raising this barrier (via coulombic, entropic or steric forces in some combination) than with reducing $\gamma$. In many cases, biliquid foams can persist for hours or days, and sometimes longer. To achieve this one must suppress not only the rupture of thin films but also the Ostwald process which, despite the more complicated geometry, still drives diffusion of A from small (few-sided) to large (many-sided) polyhedral droplets \cite{webster2,weaire}. Inclusion of a trapped species helps, as described previously, but this must now be {\em extremely} insoluble in B so as to have negligible diffusion even across the thin B films present in the foam structure. So long as they remain metastable against rupture and coarsening, biliquid foams, like soap froths, are solid materials (generally amorphous, though ordered examples can be made). As such they have an elastic modulus, and also a yield stress, both of which scale as $G\sim\gamma/R$ with $R$ the mean droplet size. This is an interesting example of a solid behaviour emerging solely from the spatial organization of locally fluid components -- for even the surfactant on the interface is (normally) a fluid film. 

\section{Effectively zero tension: stable microemulsions}
\label{tensionless}
We return now to to case 2, where a sufficient level of added surfactant can reduce $\gamma$ to negligible levels for $c_s\ge\tilde c$.  This can lead to thermodynamically stable emulsions, generally called ``microemulsions". As previously described, once this happens, enough A-B interface is created to accommodate all surplus surfactant, of which the concentration is $c_s-\tilde c$. 
Since $\tilde c$ lies below the CMC, which is generally itself small, the interesting range is usually $c_s\gg \tilde c$ so that one can treat effectively all the surfactant as interfacial. The interfacial area $\S$ of the fluid film then obeys
\begin{equation}
\frac{\S}{V}  =  (c_s-\tilde c)\Sigma \simeq c_s\Sigma = \frac{\phi_s}{v_s}\Sigma \label{area}
\end{equation}
Here $\Sigma$ is a preferred area per surfactant molecule; $\phi_s$ is the volume fraction of surfactant and $v_s$ its molecular volume. Note that principle, the area per molecule could deviate from its preferred value -- an effect important when studying, for instance, Langmuir-Blodgett films of {\em insoluble} amphiphiles such as lipids.  However, the soluble surfactants  normally used for emulsification have binding energies to the interface that are only modest (say $5-12 k_BT$), so they can adsorb and desorb from the interface on a short timescale. These processes can rapidly restore the preferred area per molecule. 

\begin{figure}[hbtp]
	\centering
	\includegraphics[width=7.truecm]{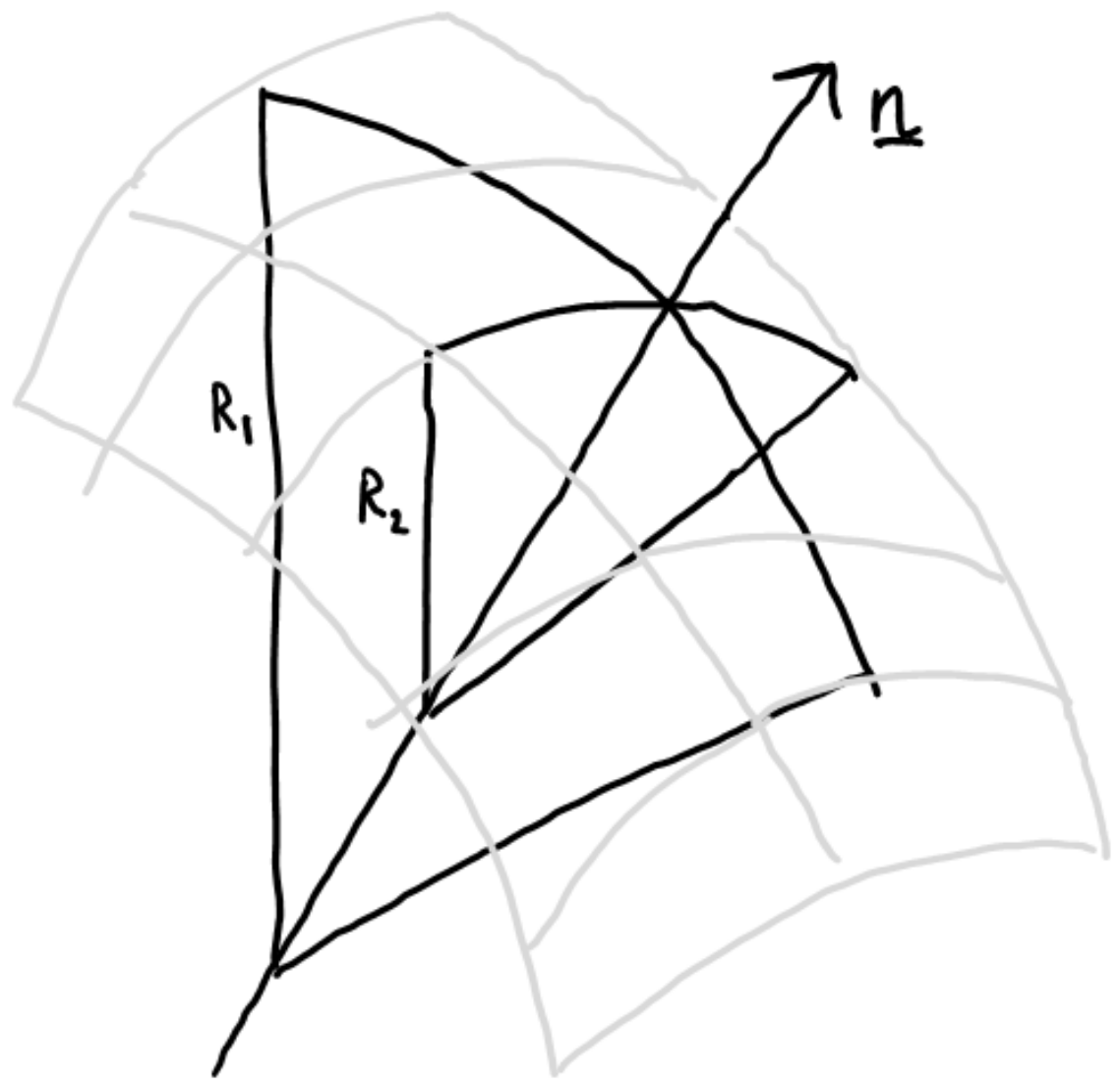}
	\caption[]{The grey lines are intended to represent a curved surface such as a piece of bicycle tyre. At the point whose surface normal is indicated by ${\bf n}$, the construction of the principle radii of curvature $R_1$ and $R_2$ is shown. }
	\label{seven}
\end{figure}

Treating $\Sigma$ as constant, the specific interfacial area $\S/V$ is fixed directly by $\phi_s$ via (\ref{area}). We next ask, what is the configuration of the interface? This is set by a competition between entropy, which prefers small, wiggly structures, and bending energy, which prefers extended, smooth ones. The bending energy can be treated by a leading order harmonic expansion about a state of preferred curvature that is set by the molecular geometry of the surfactant layer (and is tunable by varying that geometry, or by mixing surfactants with different curvature preferences). By a theorem of differential geometry, at each point on the A-B interface one can uniquely define two principal radii of curvature, $R_1, R_2$. These are the radii of two circular arcs lying in perpendicular planes to one another, each of whose centres lie on the surface normal, such that each arc kisses the surface without crossing it; see Fig.\ref{seven}. (Alternatively, $C_{1,2} = 1/R_{1,2}$ are the eigenvalues of the surface curvature tensor; see F. David's chapter in \cite{nelson}.) The radii are signed quantities and we shall take them positive for curvature towards A. For a spherical droplet of A of radius $R$, we have $R_1=R_2 = R$ whereas for a cylinder of radius $R$, we have $R_1 = R$ and $R_2 = \infty$. A saddle shape has $R_{1}$and $R_{2}$ of opposite signs. 

The harmonic bending energy then reads
\begin{equation}
F_{bend} = \int d\S \left[\frac{K}{2}\left(\frac{1}{R_1}+\frac{1}{R_1}-\frac{2}{R_0}\right)^2 +\frac{\bar K}{R_1R_2}\right]\label{bending}
\end{equation}
There are 3 material parameters, $K, \bar K$ and $R_0$. Note that a general expansion to second order in two curvatures would have five parameters ($a_1 C_1^2 +a_2 C_2^2 + a_3C_1C_2 + a_4 c_1 + a_5 c_2$) but for a fluid film $a_1 = a_2$ and $a_4 = a_5$ by rotational symmetry; the remaining expression can be reorganized to give (\ref{bending}) \cite{nelson}. The chosen parameters $K$ and $\bar K$ have dimensions of energy while $R_0$ is a length defining the preferred radius of mean curvature. Theories exist to relate these quantities to the molecular geometry of surfactants \cite{safran}, but we do not pursue these here. 

We now note that the interface between A and B cannot end in mid air: any edges must lie at the boundary of the container. (With periodic boundary conditions, no edges are possible.) The interface $\S$ can have disconnected parts (droplets) but must be orientable so that A is enclosed by it and B excluded. Moreover, the volume $V_{in}$ enclosed by the interface obeys
\begin{equation}
\frac{V_{in}}{V} = \Phi_A+\frac{\phi_s}{2} \equiv \Phi \label{Psi}
\end{equation}
Here we have partitioned the surfactant equally between A and B to allow us to define the volume $V_{in}$ as enclosed by a mathematical surface of no thickness; the phase volume of $V_{in}$ is then $\Phi$ (with $V_{out} = 1-\Phi$), and a completely symmetric state has $\Phi = 1/2$.   

To examine the statistics of the interface and determine its free energy, we now ought to compute $F = -k_BT\ln Z$ where
the partition function
\begin{equation}
Z = \int \exp[-\beta F_{bend}]{\mathcal{DS}}
\end{equation}
is found by integrating the Boltzmann weight computed using (\ref{bending}) over all surfaces $\S$ that enclose volume $\Phi V$. This has been an active area of statistical mechanics for over 30 years, and we have time here only for a brief flavour of the topic. Approaches range from crude estimates (exemplified below) to what is essentially string field theory. Although our problem is not quantum mechanical, it shares with string theory the thorny issue of how exactly to count the continuum of distinct configurations accessible to a two dimensional fluid manifold embedded in a higher dimensional space (here) or space-time (strings) (see F. David in \cite{nelson}).

\section{Some useful concepts relating to bending energy}
Here we summarize three useful concepts, using which large parts of the statistical physics problem referred to above can be qualitatively understood.

\subsection{Gauss-Bonnet theorem}
This theorem states that
\begin{equation}
\int \frac{1}{R_1R_2}d\S = 4\pi[N_c-N_h]
\end{equation} 
here $N_c$ is the number of components of our surface (where a component is a disconnected piece such as a droplet) and $N_h$ is the number of handles. A handle is a doughnut-like connection between one part of the surface and another. Thus for a sphere $N_c = 1$ and $N_h = 0$ whereas for a torus, $N_c = 1$ and $N_h=1$. Accordingly the bending energy term governed by $\bar K$ in (\ref{bending}) vanishes for a torus but not a sphere. Importantly, the result is topologically invariant so that any deformation of a torus still has zero for this quantity and any closed droplet that  is deformable continuously into a sphere has the same value of $4\pi\bar K$. Thus the bending constant $\bar K$ contributes a term to $F_{bend}$ that does not care about the local deformations of the surface, only its topology.

\begin{figure}[hbtp]
	\centering
	\includegraphics[width=11.truecm]{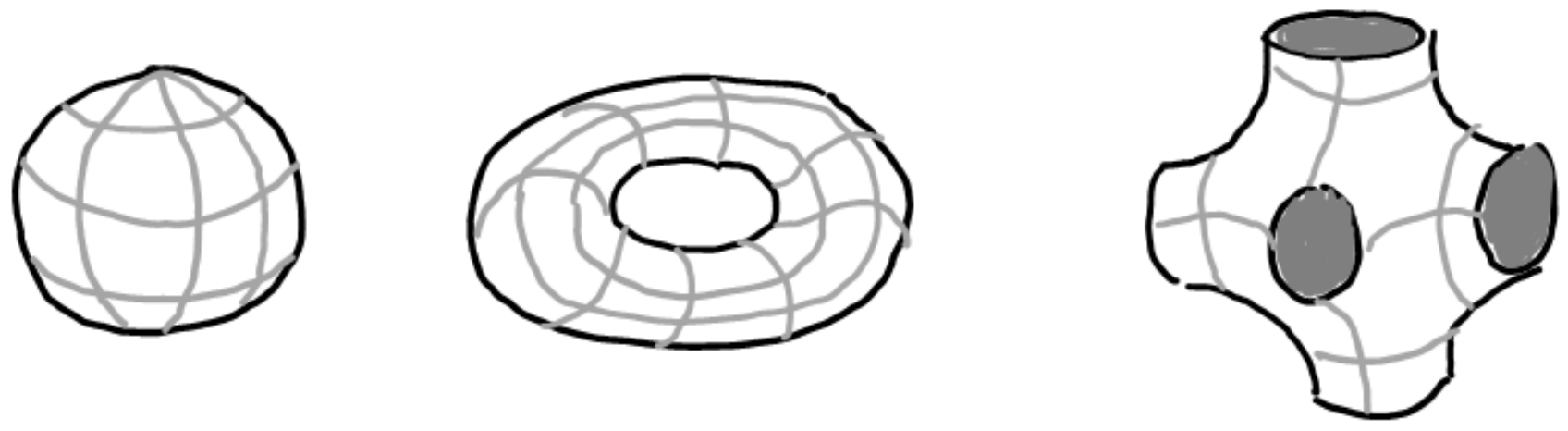}
	\caption[]{A sphere, a torus, and the unit cell of a periodic surface of constant (approximately zero) mean curvature. The hole through the torus is a handle. The grey discs on the periodic surface are cuts across it at the junction points between unit cells. Gluing a pair of these discs together at the faces of the unit cell creates one handle. Thus the final periodic structure has three handles per unit cell, but only one global component (since the entire surface becomes a single connected object). }
	\label{eight}
\end{figure}

In surfactants that stabilize emulsification, $\bar K$ is generally negative. To understand this, one needs to be aware of the existence of periodic surfaces of constant mean curvature. These comprise a periodic surface element (Fig.\ref{eight}) which connects with identical copies of itself in neighbouring unit cells to create a structure with only one global component, but several handles per unit cell. Such surfaces can be made entirely of saddles having the required mean curvature $R_0$ (at least when $R_0$ is large so the preferred curvature is weak or negligible). Accordingly the $K$ term in the bending free energy (\ref{bending}) vanishes everywhere. Because the $K$ term vanishes, all depends on $\bar K$. If this is positive, handles are favoured, and this means that the periodic structure would like as small a unit cell as possible. (Its shrinkage is ultimately controlled by anharmonic terms in the free energy that were not included in (\ref{bending}).)  The result is a short length-scale triply periodic liquid crystal, which is quite an interesting structure in itself \cite{hyde} but not an emulsion as such. Thus we assume $\bar K \le 0$ in what follows. We shall also assume $2K+\bar K >0$ so that the bending energy of a sphere is positive (otherwise one expects instead a proliferation of tiny spheres). 

\subsection{Persistence length}
We now set $R_0 = \infty$ so a flat interface is preferred. The bending energy can then be evaluated for small fluctuations in shape described by a height field $h(x,y)$ above a flat reference plane. One finds \cite{helfrich}
\begin{equation}
F_{bend} = \int \left(\frac{K}{2}(\nabla^2h)^2\right) dx\,dy \simeq \frac{K}{2}\sum_q q^4|h_q|^2 \label{monge}
\end{equation}
where in the first expression $\nabla^2$ is defined with respect to the $x$ and $y$ coordinates and in the second we have taken a fourier transform of the height field. 

Equipartition of energy then demands
\begin{equation}
\langle|h_q|^2\rangle \propto \frac{k_BT}{Kq^4}
\label{spectrum}\end{equation}
which holds for $q\le \pi/\ell$ with $\ell$ some cutoff length comparable to the thickness of the surfactant film. 
From this it is a simple exercise to show that
\begin{equation}
\langle|\nabla h(r)-\nabla h(0)|^2\rangle \propto \frac{k_BT}{2\pi K}\ln\left(\frac{r}{\ell}\right)
\end{equation}
Thus the orientation of the surface deviates from its initial value with a logarithmic dependence on separation; when this deviation is large, the expansion underlying (\ref{monge}) breaks down. An equivalent statement is to introduce a persistence length \cite{taupin}
\begin{equation}
\xi_K \simeq \ell \exp\left[\frac{4\pi K}{\alpha k_BT}\right]
\label{xiK}\end{equation} 
where $\alpha$ is sometimes referred to as a `geometrical' constant. The interpretation is that a fluid interface governed by (\ref{bending}) completely loses its sense of orientation at length scales beyond $\xi_K$. 

Because of the exponential factor, $\xi_K$ is exquisitely dependent on $K$ (and also on $\alpha$, which turns out to obey $\alpha = 3$ as discussed below). Thus for $\ell = 1$ nm, $\xi_K \simeq 1\mu$m when $K = 1.65k_BT$. For $K/k_BT = 3$, we already have $\xi_k\ge 300\mu$m, and $\xi_K$ is irrelevantly large, for our purposes, once $K/k_BT$ is much larger than this. (Indeed for $K/k_BT = 6$ we have $\xi_K = 9$cm.)

\subsection{Renormalization of the bending constant} 
In many cases we do not want to describe the A-B interface in complete detail, but instead need a coarse grained description of its properties on some geometric length scale $\xi$ set by, for instance, the size of droplets in an emulsified state. Under coarse graining we replace an entropically wiggly interface, with structure at short scales characterized by the bending spectrum (\ref{spectrum}), by a smooth one on the scale $\xi$ for which such structural detail is still present but removed from the description. To allow for that, we need to accept that the effective bending constant for the smoothed-out interface is softer than the microscopic one. Indeed the result of a careful and quite technical field-theory calculation (see F. David in \cite{nelson}) is that, to leading order, 
\begin{equation}
K_{eff} = K - \frac{3k_BT}{4\pi}\ln\left(\frac{\xi}{\ell}\right)
\label{Keff}\end{equation}
This is quite plausible in that it suggests there is effectively no resistance to macroscopic curvature at length scales beyond $\xi_K$ as defined in (\ref{xiK}). Indeed consistency with this calculation requires the choice $\alpha = 3$ as previously mentioned. However, the numerical prefactor in (\ref{Keff}), which indeed fixes this value of $\alpha$, is far from obvious and was the subject of heated debate for a number of years. 

\section{Some consequences of bending energy physics} 
In addressing this large topic, we will have time for only a few limiting cases.

\subsection{Emulsification failure}  
\begin{figure}[hbtp]
	\centering
	\includegraphics[width=11.truecm]{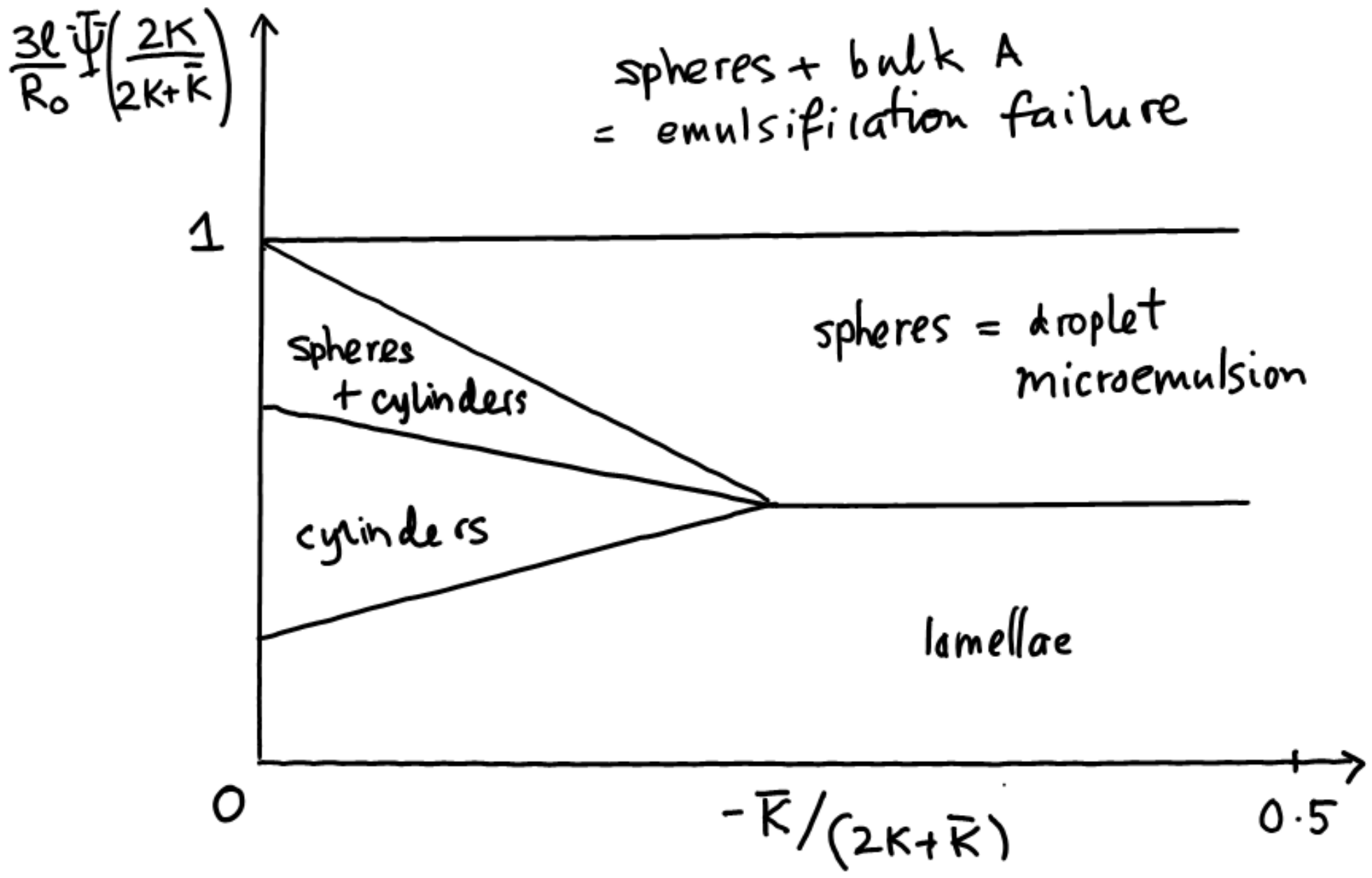}
	\caption[]{Schematic phase diagram (after \cite{safran,turk}) showing the phase of lowest free energy as a function of enclosed phase volume $\Psi$, the elasticity parameters $R_0,K,\bar K$, and the film thickness $\ell$. This is a simplified calculation which does not account for entropy nor the existence of miscibility gaps between the states shown. Nor does it include anharmonic corrections to the bending energy.}
	\label{nine}
\end{figure}

We first consider $\bar K>0$, and $\ell \ll R_0 \ll \xi_K$. This holds for interfacial films with finite preferred radius of curvature that are relatively stiff on that length scale. Entropy is negligible: we need only minimize $F_{bend}$ at fixed $\S V$. The computation of $F_{bend}$ for spheres, cylinders and lamellae is trivial and for simplicity we limit attention only to these geometries. We have
\begin{equation}
F_{bend} = 4\pi\left(2K\left[1-\frac{R^2}{R_0^2}\right]+\bar K\right)
\end{equation}
for a sphere of radius $R$; 
\begin{equation}
F_{bend} = \frac{\pi K L}{R_0}\left[1-\frac{2R}{R_0}\right]^2
\end{equation}
for a cylinder of radius $R$ and length $L$; and $F_{bend} = 2KA/R_0^2$ for a flat sheets of area $A$. 
A schematic phase diagram found by comparing these forms \cite{safran} is shown in Fig.\ref{nine}. This shows, as one might expect, a preference for lamellae when $R_0$ is large and spheres when it is small. More interestingly, it is a simple exercise to show (by equating the enclosed volume to $V\Phi$ and the surface area to $\S$) that the droplet size $R$ for (monodisperse) spheres obeys
\begin{equation}
R_s = \frac{3\Phi \ell}{\phi_s}
\end{equation}
where $\ell = v_s/\Sigma$ is the thickness of the surfactant film. (Here $v_s$ is an effective molecular volume for surfactant.)
If $R_s<R_0$, then, modulo slight shifts relating to the value of $\bar K$, the droplet phase is stable. However, if surfactant is removed or the internal phase volume fraction $\Phi$ is increased to the point where $R_s$ found above exceeds $R_0$, there is no advantage to the system in paying the additional bending cost of having droplets larger than their preferred curvature radius. Instead, the droplets remain of size $R_0$, so that the value of $\Phi$ in the droplet phase is less than its globally imposed value. This means that an excess phase comprising bulk A-rich fluid is expelled from the system. The system has created the perfect interface (in area and curvature) for the amount of surfactant present, but this fails to enclose all of the dispersed phase -- a situation known as emulsification failure \cite{safran,turk}.

\subsection{Bicontinuous microemulsions} The second case we address here is where $R_0 \gg \xi_K$; for simplicity we treat $R_0$ as infinite so the preferred interfacial state is symmetric between A and B (sometimes called a `balanced' system). We also assume $\ell \ll \xi_K \ll 100\mu$m, so that entropy matters (including the renormalization of $K$) and can compete with bending energy on roughly equal terms. 

\begin{figure}[hbtp]
	\centering
	\includegraphics[width=11.truecm]{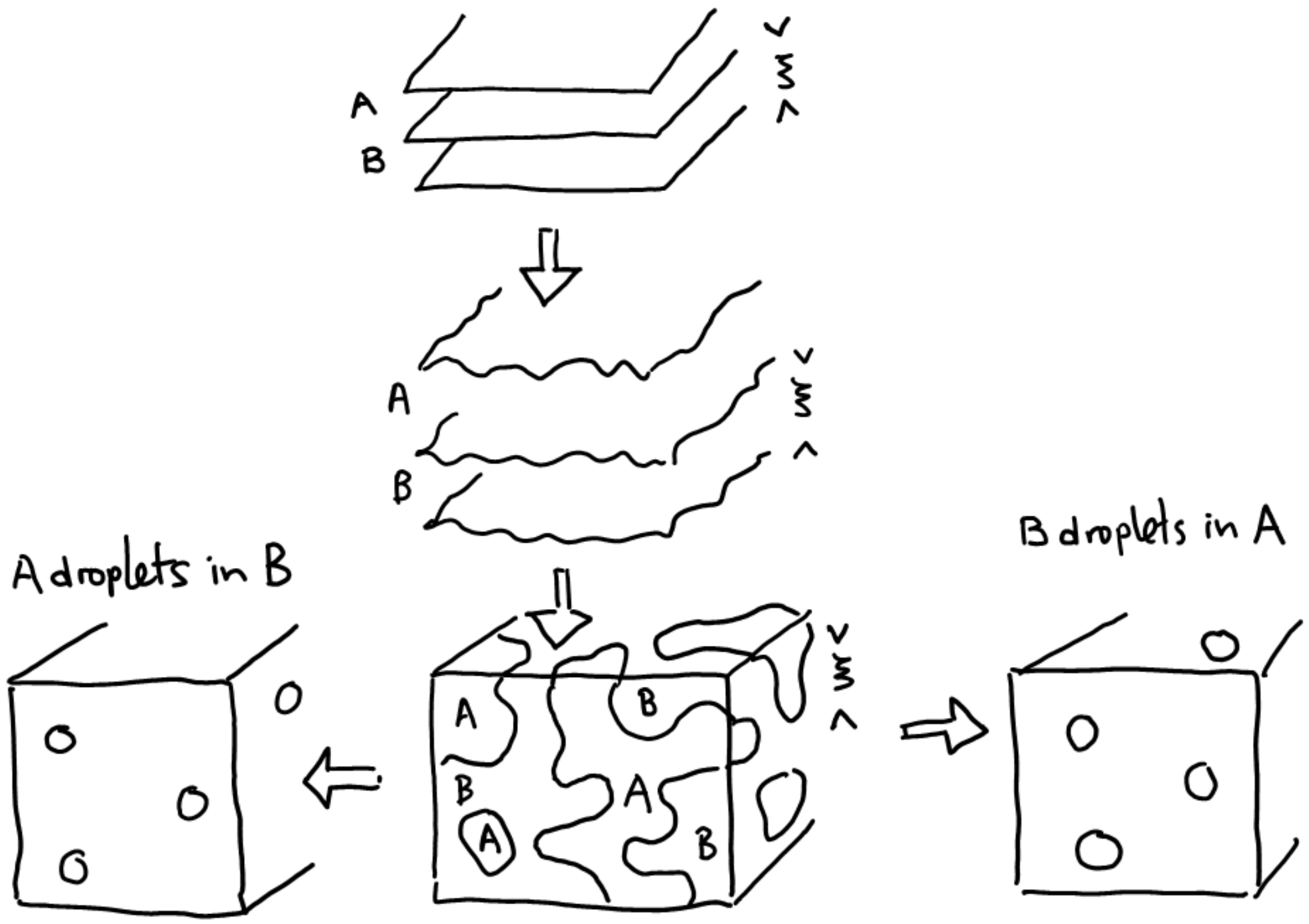}
	\caption[]{Schematic evolution from lamellae to bicontinuous microemulsion as the concentration of surfactant (and hence $\xi_K/\xi$) is decreased in a system with $\Psi\simeq 0.5$ (three central frames). The breakdown into droplet phases occurs on varying $\Psi$ away from $0.5$, with details dependent on $\bar K$.}
	\label{ninepointfive}
\end{figure}

Assuming $\Phi$ of order 0.5 (roughly symmetric amounts of A and B) we can introduce a structural length scale $\xi$ which is then set by $\phi_s$. Specifically for a lamellar phase one has
a layer spacing $\xi$ between adjacent surfactant films set by
\begin{equation}
\phi_s \simeq \frac{\ell}{\xi+\ell}
\end{equation}
(This is subject to logarithmic corrections from fluctuations which need not concern us here.)
When $\xi\simeq \ell$ the system has no option but to fill space with flat parallel layers. As $\phi_s$ is reduced ($\xi$ raised) the layer spacing $\xi$ becomes comparable to $\xi_K$. For $\xi/\xi_K \le 1/3$ (or so), the lamellar phase fluctuates but remains stable. However if $\phi_s$ is then decreased further so that $\xi \simeq \xi_K$, these layers melt into an isotropic phase comprising (for $\Phi\simeq 0.5$) bicontinuous domains of A and B fluids separated by a fluctuating surfactant film (Fig.\ref{ninepointfive}) \cite{taupin}. This is the {\em bicontinuous microemulsion} and represents a thermodynamic route to prevent coarsening of the transient bicontinuous emulsion structures encountered in Section~\ref{PSK}. If $\Phi$ now deviates strongly from $0.5$, then (just as found there) the structure depercolates, forming a droplet phase; the details of where and how this happens involve $\bar K$ which, we recall, is sensitive to topological changes. The droplet phase so formed is somewhat different from the one discussed above at $R_0\ll\xi_K$, since this one is stabilized by entropy and fluctuations, not by a preferred curvature of the droplets. 

Several theories of the bicontinous microemulsion were developed in the 1980s. Some of the most successful \cite{us,lulu} used coarse grained lattice models in which fluid domains are placed at random on a lattice of some scale $\xi$; the bending energy and area of the resulting interface can be estimated and used to calculate a phase diagram. One specific feature is the appearance of three-phase coexistence in which a  `middle phase' microemulsion coexists with excess phases of both oil and water. This is somewhat analogous to a double-sided emulsification failure in which, for a fixed quantity of surfactant, the system creates a happy interfacial structure on a certain length scale and then rejects any excess oil and/or water. This length scale is now set by $\xi_K$ rather than $R_0$. Although the models do predict this, it remains somewhat unclear even now why this structure does not want to fragment further. Such fragmentation would lead to two coexisting phases of dilute A droplets in B and vice versa. One possibility is that $\bar K_{eff}(\xi)$, whose renormalization properties we have not discussed (see F. David in \cite{nelson}), becomes positive at $\xi\ge \xi_K$. This would cause condensation into a handle-rich state, similar to the periodic ones previously stated to arise for $\bar K >0$, but stabilized now by an entropy-driven tendency to form handles that only operates at large enough length scales.

\subsection{The sponge phase} One final twist on this story concerns systems in which there is a huge phase volume aysmmetry between A-rich and B-rich fluids, but where the surfactant has a very strong molecular preference to form a flat film rather than highly curved structures such as micelles. The interfacial structure that forms spontaneously at $\gamma \simeq 0$ is, in the almost complete absence of B, necessarily now a bilayer with A (usually water) on both sides and a thin B layer in the middle. It is easy to imagine the lamellar state, and for small volume fractions of bilayer this suffers the same instability towards melting as described above for the microemulsion. The result is a bilayer film that now separates two randomly interpenetrating domains containing the same solvent A: this is usually called the `sponge phase'. Perhaps surprisingly, the quantity $\Phi$ which discriminates between $V_{in}$ and $V_{out}$ remains meaningful. However it is now fixed not by the global phase volumes of the two solvent domains (since these contain the same solvent and are interchangeable) but instead takes whatever value minimizes the free energy. Moreover, defining $\Psi = 2\Phi-1$, the microscopic free energy of this system is invariant under $\Psi\leftrightarrow -\Psi$ (an operation which exchanges the identities of the `inside' and `outside' fluid domains). This symmetry can remain intact in the physical state of the system, or it can break spontaneously, for instance by the formation of a phase of closed spherical vesicles \cite{epl,leibler}. The symmetry breaking can be continuous, and then shows critical behaviour of Ising-like character and a unique scattering signature \cite{sponge}.

This scenario is theoretically quite rich and elegant \cite{coulon}, but the sponge phase contains at most a few percent of its second solvent (B, as described above) sequestered within the bilayers. It is therefore not particularly useful for emulsification and has found relatively few applications in technology so far. This contrasts somewhat with our next topic.

\chapter{Particle-Stabilized Emulsions}\label{PE}
A typical soluble surfactant has a hydrocarbon tail containing 10-20 repeat units and a head group with either a dissociable salt (ionic or anionic surfactants) or a polar polymer such as a string of ethoxy groups (typically 4-8 of these in a nonionic surfactant). Such surfactant molecules generally have energies of attachment to the oil-water interface of between $5k_BT$ and $20k_BT$, high enough to alter interfacial properties but low enough to allow rapid adsorption and desorption to maintain local equilibrium. However, it is easy to create amphiphilic species that are much larger \cite{ettelaie,lips}; these range from lipids, via block copolymers (tens to tens of thousands of repeat units) and globular proteins, to so-called `janus beads'. The latter are colloidal spheres, up to a micron in size, with surface chemistry that favours water on one hemisphere and oil on the other. For typical solid-fluid interfacial tensions ($\simeq 0.01$ Nm$^{-2}$) janus beads have attachment energies of order $10^7k_BT$ or larger. Such species are adsorbed irreversibly in the sense that Brownian motion will never lead to detachment. All of these amphiphilic particles can, in principle, lead to a vanishing of the thermodynamic interfacial tension if micellization and similar processes are avoided. However, thermodynamic concepts such as interfacial tension itself turn out to be of very limited value in understanding irreversibly adsorbed layers.

\section{Adsorption of non-amphiphilic colloids}
In fact, although janus particles can be made and are often studied \cite{janus}, they are rarely used in practical applications to stabilize emulsions. This is because a simpler route exists, in which these purpose-built amphiphilic colloids are replaced by those of homogeneous surface chemistry: that is, ordinary colloidal spheres \cite{binks,binks2,aveyard}. 

Such particles can also have attachment energies to the interface that are vastly larger than $k_BT$. This is most easily seen when the surface chemistry is chosen with equal affinity towards water or oil so that the two solid--fluid interfacial tensions, $\gamma_{SO}$ and $\gamma_{SW}$ are the same. The energy of such a  particle (of radius $a$) is independent of where it resides in an oil-water system, but the energy of the oil water interface is reduced by $\pi a^2 \gamma$ if the particle is placed there. This is simply the surface energy of the interfacial patch now covered by the colloid, and $\gamma$ is the usual fluid-fluid interfacial tension. Setting $\gamma = 0.01$Nm$^{-2}$ gives $\Delta F_{attach} \simeq 10^7k_BT$ for $a=1\mu$m and $\Delta F_{attach} \simeq 10k_BT$ for $a=1$nm. Thermal detachment thus remains negligible for $a\ge 5$nm.

\begin{figure}[hbtp]
	\centering
	\includegraphics[width=7.truecm]{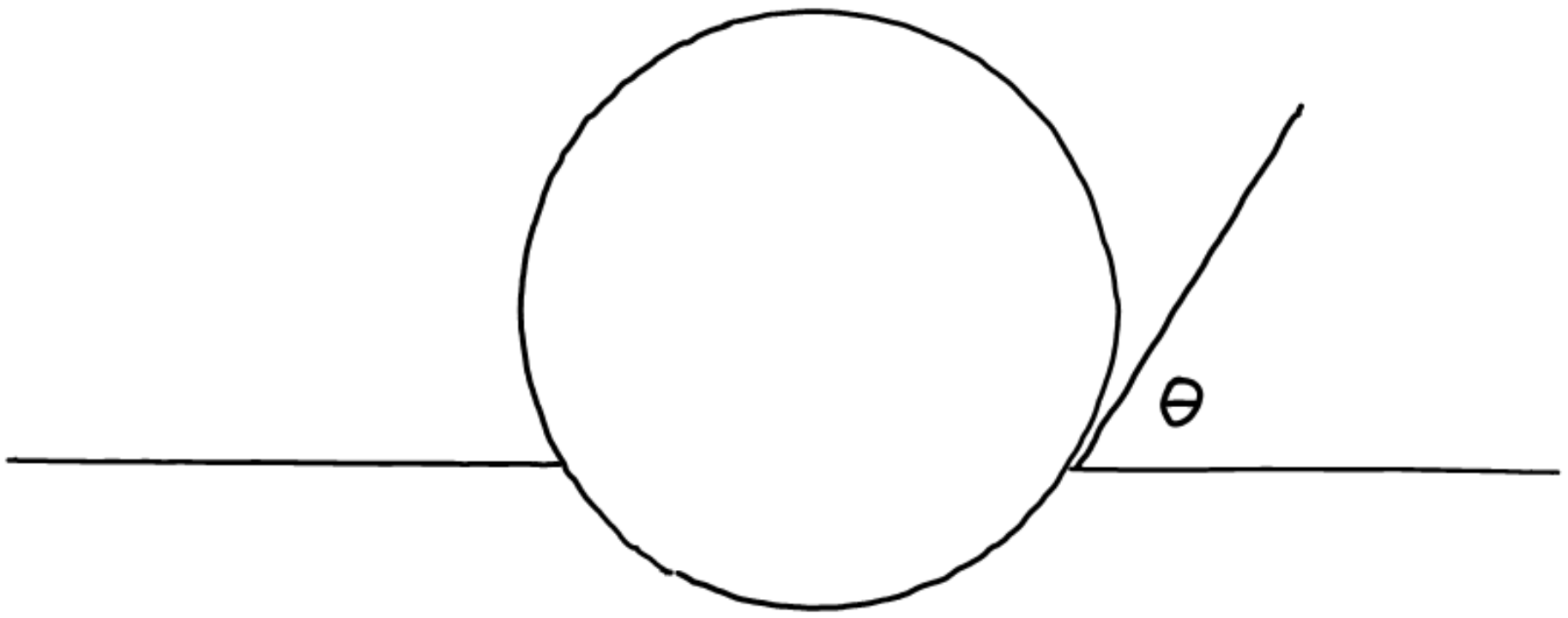}
	\caption[]{Geometry of a partially wet colloidal particle at the fluid fluid interface; $\theta$ is the contact angle. This is usually measured through the polar phase, so the upper fluid is water and the lower oil as drawn here.}
	\label{ten}
\end{figure}

More generally, the two solid-fluid tensions are different. However, the lowest free energy state has the colloid on the interface so long as the contact angle $\theta$, which obeys
\begin{equation}
\gamma \cos \theta = \gamma_{SO}-\gamma_{SW}
\end{equation}
obeys $0\le\theta\le\pi$. (This is the definition of partial, as opposed to complete, wetting.) In the absence of a body force such as gravity acting on the particle, the interface remains perfectly flat; the particle is displaced so that it intersects the interface at angle $\theta$ (Fig.\ref{ten}). Accordingly the area of the covered disc is reduced; the attachment energy is also, but remains very large compared to $k_BT$, at least for $a \ge 10$nm, unless $\cos\theta$ is very close to $\pm 1$. Indeed, a calculation of the force $f_D$ required to detach the particle gives \cite{binks}
\begin{equation}
f_D = \pi a \gamma (1\pm |\cos\theta|) \label{detach}
\end{equation} 
with the $+(-)$ sign applies when the particle is being pulled out of (into) its preferred solvent. Modest deviations from the neutral wetting condition ($\cos\theta  = 0$) thus have little effect on the detachment barrier.

Turning now to the case of a layer of adsorbed particles, it is often said that deviations from neutral wetting creates spontaneous curvature, as was described previously for surfactants. (In principle this effect looks very strong, with preferred curvature radii of order $a/\cos\theta$.) However, this is deceptive, since the particles are perfectly happy on a flat interface unless their density is high enough to jam them into contact. Beyond this point, it is true that the interfacial area can be reduced further at fixed particle coverage by introducing curvature, but without any constraint on the overall interfacial geometry it is even better to expel the particles and have no interface at all. Usually the overall geometry is constrained purely by the necessity of having an interface between bulk coexisting phases; in that case, regardless of $\theta$, the lowest free energy state is found first by creating a flat interface, and then by covering as much of it as possible with particles. At no point is a curved interface actually preferred.

\section{Particles on curved interfaces}

Consider now placing a number of colloidal particles on a spherical droplet of fluid A in fluid B, or vice versa. It is crucial to note that each colloid can be accommodated with the required contact angle $\theta$ simply by cutting a small spherical cap out of the interface and slotting the particle into place there. The contact line is a perfect circle, as required for tangency at fixed angle to a sphere. To conserve the enclosed volume the droplet radius may change slightly, but it remains perfectly spherical.
There is no change in Laplace pressure, and the energy is completely independent of {\em where} the colloids are placed. Accordingly there is no (tension-mediated) capillary force between the particles. These statements can only change if particles become jammed so they interact directly via particle-particle forces. 

Contrast this with the case of spherical colloids on a hypothetical cylinder of fluid. It is not possible now to insert a spherical particle into the surface of this cylinder at fixed contact angle $\theta$ unless the fluid interface becomes deformed. This deformation costs extra interfacial energy, and can be minimized by placing two particles close together rather than far apart. Accordingly, there is a capillary attraction between the colloids, which will have a strong tendency to aggregate. Similar arguments apply to nonspherical particles such as ellipsoids, even on flat surfaces \cite{stebe,lehle}. More generally, this field offers various formal mathematical problems involving area minimization under the constraints of enclosed volume and adsorbed particle number, many of which seemingly remain unexplored. \label{classical}

\section{Particle-stabilized emulsions}
The use of partially wetting colloidal particles to stabilize emulsions dates back at least one century \cite{binks2}. Unless janus particles are used \cite{aveyardSM}, the resulting `Pickering emulsions' are {\em always metastable}: as stated already, the minimum free energy state comprises bulk A and B phases, separated by a flat interface, with as much of this interface as possible covered by particles. (This requires an ordered hexagonal packing in principle.) The remaining particles are then distributed randomly in their preferred solvent (if $\theta \neq \pi/2$) or in both solvents if there is no preference ($\theta = \pi/2$). 

However if the system is stirred or otherwise agitated, it is quite easy to create robust droplets, whose size is fixed by the production process as well as the fluid phase volumes and the particle density. Closely related droplet phases can also be made in which the dispersed phase comprises air or vapour rather than liquid \cite{stone}. 

\subsection{Resistance to coalescence}
One specific route to Pickering emulsions \cite{binks,arditty} is to create a vast number of very small droplets by applying extreme flow conditions to temporarily mix fluids A and B. The flow sweeps particles onto the interface whose initial surface area is, however, much larger than they can cover. Coalescence initially proceeds as normal (assisted by maintaining a lower but finite flow rate).  During a coalescence between two droplets of size $R$, the total volume they enclose is conserved, so the radius changes as $R\to 2^{1/3}R$ and the surface area as $\S \to 2^{-1/3}\S$. Thus the area fraction of particles increases by $2^{1/3}$ each coalescence until the surface density is high enough to prevent further coalescence. 

Because it requires significant local rearrangement of the particles in relation to the interface, the resulting energy barrier scales as $\pi a^2\gamma \gg k_BT$. (Indeed, Pickering formulations 
often offer exceptional stability against coalescence compared to conventional surfactants.)
This requires a coverage of comparable to, but significantly below, that of a densely packed 2D amorphous film.

In most cases the protected droplets therefore have a fluid layer of particles at their surface not a jammed one, so that they relax to a spherical shape. In some cases, though, two droplets coalesce that are already nearly protected. The reduced surface area of the combined droplet, if spherical, is then not enough to accommodate all the particles. Coalescence proceeds, but then is arrested in a jammed state comprising a non-spherical droplet, whose surface particles are clamped in position by interfacial tension \cite{clegg}.  (Related effects can be seen at lower particle coverage if there are strong attractive forces between colloids creating a bonded rather than a jammed interfacial layer.) Similar jammed structures can again been seen in armoured air bubbles \cite{stone}.

\subsection{Resistance to Ostwald ripening}
Recall that a close packed monolayer of particles (whether ordered or amorphous) can be placed on the surface of a spherical fluid droplet of radius $R$ (say) without altering its interfacial geometry. Imagine such a droplet with the particles just in contact with one another. There is still a fluid-fluid interface at the interstices between particles, and this has the same curvature, and hence Laplace pressure $\gamma/R$, as the original drop. 

\begin{figure}[hbtp]
	\centering
	\includegraphics[width=7.truecm]{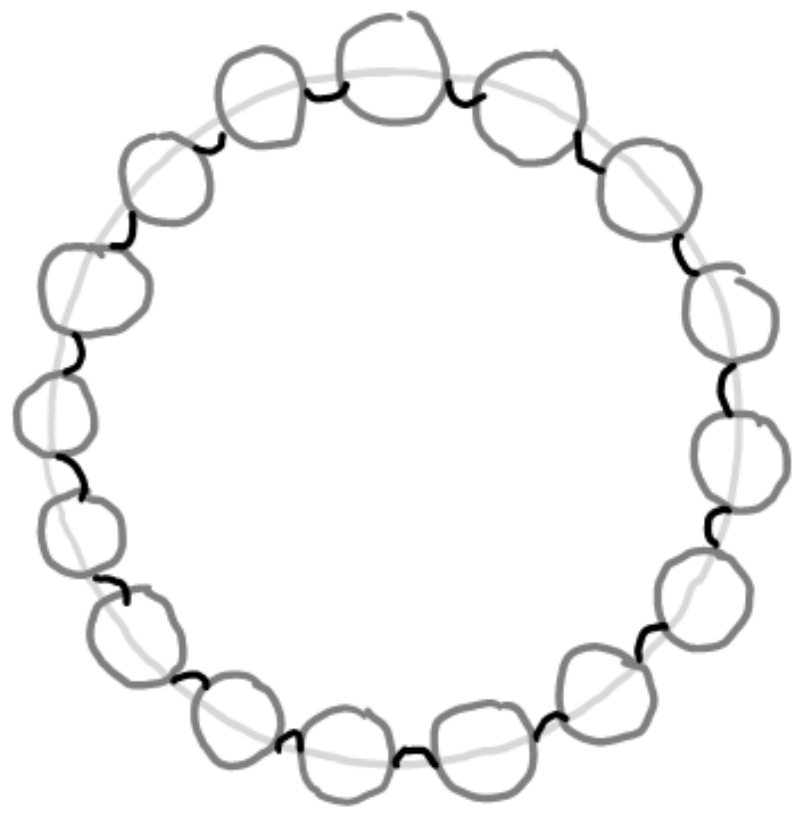}
	\caption[]{In light grey is the initial locus of a fluid interfaces into which are inserted the particles shown in darker grey. These are jammed in a 2D layer but have interstitial fluid regions as shown between them. If the volume of fluid in the droplet is now reduced, to maintain a fixed contact angle with particles that cannot move, the curvature of the interface is reversed to give the final fluid locus (black). This creates a negative Laplace pressure which switches off the Ostwald process.}
	\label{eleven}
\end{figure}

Suppose now that this droplet is in diffusive equilibrium with one or more larger ones. According to the Ostwald mechanism, it will start to shrink. However, if the particles are already in contact they cannot follow the droplet surface inwards as this happens. Moreover, each particle demands an unchanged contact angle with the interface, which is effectively now pinned to the particle layer. It is easy to see (Fig. \ref{eleven}) that even a small loss of volume of the droplet under these conditions will cause an elimination of the Laplace pressure followed by its sign reversal. Thus a state of finite droplet radius, but zero mean curvature of the interface so that $\Delta P = 0$, presumably exists. (Proof of this is one of the seemingly unsolved problems referred to in Section \ref{classical} above.) The resulting droplet is fully resistant to Ostwald ripening, and therefore a suspension of Pickering droplets is also stable against it, so long all the interfacial particle layers are jammed. The mechanism is similar to that described in Section \ref{ts} using a trapped species, with the interfacial particles themselves playing that role.

If the jamming is not complete initially, as would typically be true for emulsions made by the arrested coalescence method outlined above, a minority of droplets may grow, ultimately forming a coexisting bulk phase of the dispersed fluid. However, if this is now removed, the emulsion component is highly stable against both coalescence and Ostwald ripening. There are subtleties with this simple picture, though, if the dispersed phase contains more than one soluble component \cite{compripe}.  

\section{Particle detachment}
The barrier to individual particle detachment ceases to be very large compared to $k_BT$ for nanometre scale particles. (With such particles, the interfacial energies we have described may need supplementing with an energy term proportional to the length of the contact line \cite{aveyard}.) The same can apply for larger colloids if the interfacial tension $\gamma$ is relatively small, or if surfactants are added that change the particles' wettability. Low tension often arises for polymer polymer mixtures as opposed to oil and water; and blending of polymers into emulsions is an important area of technology. Moreover, once particles are jammed by interfacial tension into a dense layer, the detachment is a collective process and it is not clear whether some pathways then exist that allow particles to escape without crossing a high barrier \cite{kim1}. 
Addressing the physics of Pickering emulsion with non-negligible detachment processes is a complicated but interesting open area.

\section{Some interesting particle-stabilized structures}
\begin{figure}[hbtp]
	\centering
	\includegraphics[width=12.truecm]{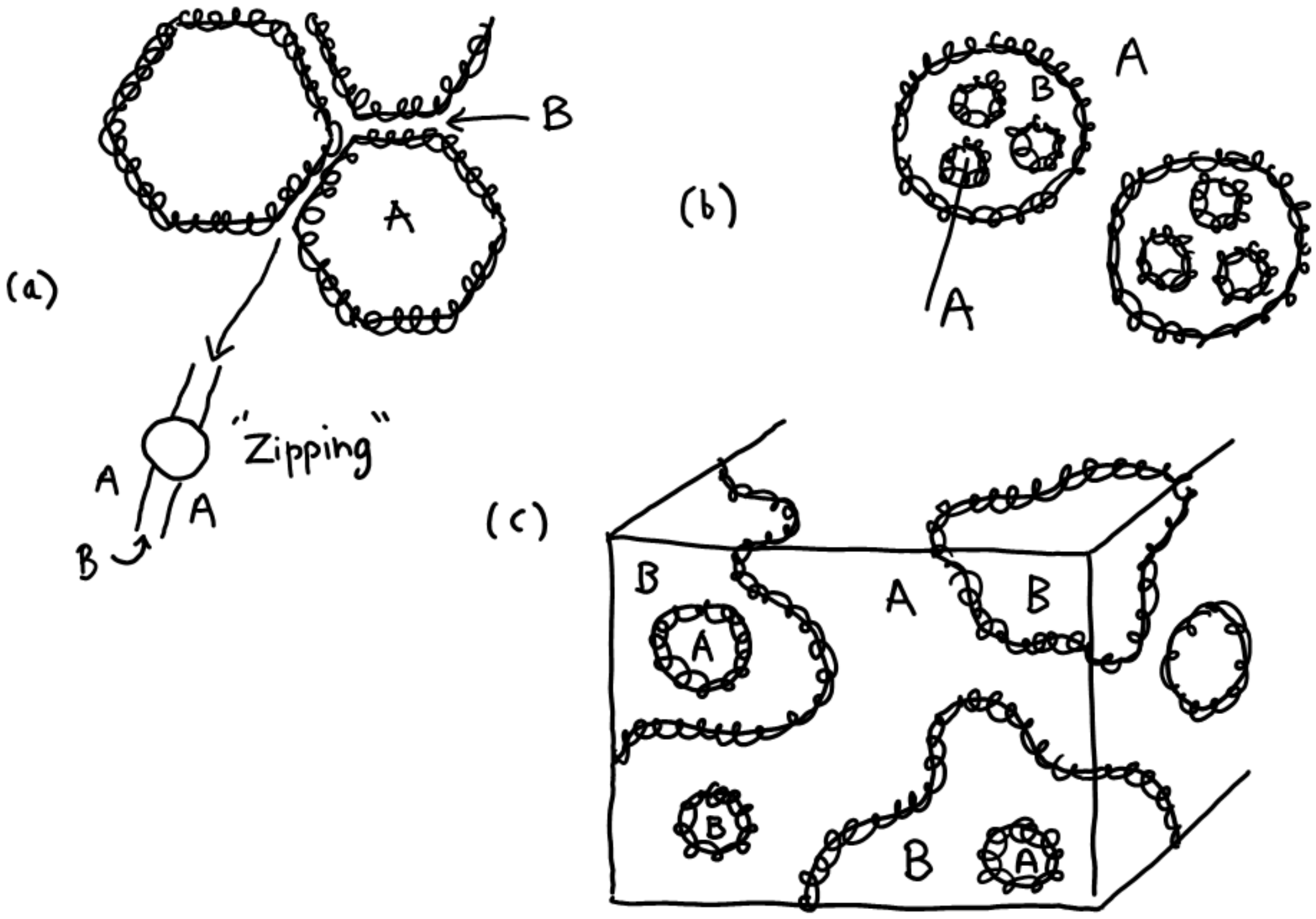}
	\caption[]{Various particle-stabilized interfacial structures between fluids A and B. Notation: the continuous squiggly line stands for a monolayer of colloidal particles. (a): A biliquid foam with thin B films separating polyhedral A droplets. Inset: the possibility of `zipping'. (b) A multiple emulsion. (c) A bijel; compare the bicontinuous microemulsion in Fig. \ref{ninepointfive}. }
	\label{twelve}
\end{figure}

More generally, there are large gaps in our theoretical understanding of particle-stabilized emulsion formulations, in part due to the fact that non-desorption causes almost all the interesting structures to be far from equilibrium.  In what follows we describe a selection of these structures, as depicted in Fig.\ref{twelve}. In at least one of these cases, simulations have been a guide to the creation of new states \cite{kevin}.

\subsection{Pickering foams}
Drainage or centrifugation of a Pickering emulsion can lead to a compressed foam structure that is like the biliquid foams discussed previously \cite{studart}. These can be quite stable thanks to the combined resistance to coalescence and Ostwald ripening provided by the particles. However, removal of the continuous phase causes droplets to deform into polyhedral shapes which increases their area and can bring the particle coverage below the threshold for protection. Interestingly, for $\cos \theta \neq 0$ it is possible for a particle to bridge across a thin film of its preferred solvent while maintaining the equilibrium contact angle at both interfaces. Though not as effective as two monolayers, the resulting `zipped interfaces' still present a barrier to film rupture \cite{zips}. If present, these might also resist re-expansion of a foam after centrifugation has ended.

\subsection{Multiple emulsions} 
Simple manual agitation of binary immiscible solvents containing partially wettable particles often results in droplet-within-droplet structures known as multiple emulsions. Such structures require stability against both coalescence and ripening for A droplets in B and B droplets in A simultaneously. This is relatively difficult for surfactant formulations but seemingly quite easy with particle-stabilized ones.

\subsection{Bijels}
Bijels (\underline{b}icontinuous \underline{i}nterfacially \underline{j}ammed \underline{e}mulsion ge\underline{l}s) are metastable analogues of the bicontinuous microemulsion: a particle layer resides at the interface between interpenetrating domains of A and B. This structure was predicted first computationally in 2005 \cite{kevin} and confirmed in the laboratory in 2007 \cite{eva}. It has no known counterpart among metastable surfactant-stabilized formulations, presumably because surfactant desorbs too easily ever to prevent the coarsening of an interface with finite $\gamma$. In a bijel, the interfacial film of non-detachable particles is clamped by tension into a 2D jammed layer, which imparts solidity to the whole 3D structure \cite{joe,jobAFM}. This robustness can be improved further by having an interaction potential between particles with a steep barrier and then an attractive minimum at short distances. The interfacial tension pushes particles over the barrier creating a permanent interfacial film \cite{whyte}.

One recipe for making a bijel is to choose a fluid pair A+B that are miscible at high (or perhaps low) temperature \cite{cleggrev}. The colloidal particles are then dispersed within the single-phase AB mixture. On quenching the temperature, the fluids separate and particles are swept onto the interface. The coarsening process arrests when a jammed monolayer is formed, creating the bijel. The final structural domain size obeys $\xi\simeq a/\phi_p$ with $a,\phi_p$ the particle size and volume fraction, and the elastic modulus of the solid bijel scales as $G\sim\gamma/\xi$. Bijels are currently being explored for various applications including materials templating \cite{mohraz}, where their A-B bicontinuity, and in some cases tricontinuity (allowing also for the percolating particle layer) could prove advantageous. 

\section{Stability against gravity}
I choose as a final topic something representative of very recent research done in my own group at Edinburgh (with experiments led by Paul Clegg). Consider a droplet of oil in water (which generally will float) stabilized by particles of higher mass density than either fluid. Each particle feels the force of gravity: for small particles this is very weak (though it can be increased by centrifugation). As we will see, the force can accumulate across particles to give much bigger effects than expected for a single particle. The same applies when particles feel body forces that are not gravitational in origin -- for example if emulsion stability is provided by magnetic colloids which are then subjected to a field gradient \cite{melle}.

\subsection{Critical Bond number}
As shown previously in (\ref{detach}), a force $f_D$ is required to remove one particle of radius $a$ from a flat interface. The same force is needed, to a good approximation, for detachment from a curved droplet of radius $R$, so long as $R\gg a$. Equating $f_D$ to the force of gravity $f_g$ gives a critical value of Bo$^*(1) = 1$ for the Bond number which (absorbing a factor 4/3 often found in the literature) we define as
\begin{equation}
\mbox{\rm Bo} =\frac{4 a^2 g \Delta \rho}{3\gamma(1-\cos\theta)} 
\end{equation}
Here $g$ is gravity and $\Delta\rho$ the density mismatch.

Our task is to generalize this result to find the critical Bond number Bo$^*(N)$ for detachment from a droplet coated with a monolayer of $N$ particles. (This will depend on the exact coverage, which we assume to be constant and fairly close to a packed monolayer.)
Because the gravitational force can be transmitted from one particle to those below that support it, it is a reasonable expectation that Bo$^*(N)<1$. If so, how does it depend on $N$? Two possible mechanisms come to mind, as follows.

\subsection{Keystone mode versus tectonic mode}
The first candidate is that a finite fraction of all the weight of the particles bears down on  a `keystone' particle at or near the bottom of the droplet. The net gravitational force is $N$ times that on one particle, while $f_D$ is effectively unchanged. The keystone particle detaches once $Nf_g\ge f_D$; this implies that Bo$^*(N)\propto N^{-1}$. A stream of detachments will continue until the number of remaining particles falls below $f_D/f_g$. This can leave the upper part of the droplet unprotected against coalescence, resulting in macroscopic instability of the emulsion.

The second candidate mechanism is that a finite fraction of the droplet breaks off in a collective `plate tectonic' detachment event. A crude estimate of when this might happen is found by balancing the gravitational force on the lower half of the droplet $Nf_g/2$, against the interfacial tension across the equator, $2\pi\gamma R$. Since $R\simeq aN^{1/2}$ this gives (within prefactors that depend on geometric details such as the deformed droplet shape) $Nf_g \simeq N^{1/2}f_D$, so that Bo$^*(N)\propto N^{-1/2}$.

For droplets containing tens to thousands of particles, the competition between these two modes seems to be finely balanced. Experiments on particles with significantly non-neutral contact angles $\theta$ suggest the keystone mode \cite{key} whereas simulations with $\theta = \pi/2$ suggest the tectonic one \cite{plate}. So a provisional conclusion is that the result depends on contact angle; this is plausible because the one-particle detachment energy (at given particle size) depends strongly on contact angle, whereas the force balance in the tectonic mode does not.

\section*{Acknowledgements}
I thank the numerous colleagues who have collaborated with me on these topics over the years, and thank the participants of the les Houches School whose lucid questioning has improved these notes in many places. I also thank Anne Pawsey, Job Thijssen and Paul Clegg for comments on the manuscript. I acknowledge funding from the Royal Society in the form of a Research Professorship, and funding from EPSRC under Grant EP/J007404. 


\end{document}